\documentclass[twocolumn]{aastex63}
\usepackage{graphicx}
\graphicspath{{./Images/}}
\usepackage{booktabs}
\usepackage{url}
\usepackage{natbib}
\usepackage{appendix}
\usepackage{subfigure}
\usepackage{longtable}
\usepackage{amsmath}
\usepackage[figuresright]{rotating}

\begin{document}

\correspondingauthor{Jared Johnson} 
\email{jjrod118@uw.edu}

\author[0000-0002-2630-9490]{Jared R. Johnson}
\affil{Department of Astronomy, Box 351580, 
University of Washington, Seattle, WA 98195, USA}
\author[0000-0001-5530-2872]{Brad Koplitz}
\affil{Department of Astronomy, Box 351580, 
University of Washington, Seattle, WA 98195, USA}
\author[0000-0002-7502-0597]{Benjamin F.\ Williams}
\affil{Department of Astronomy, Box 351580, 
University of Washington, Seattle, WA 98195, USA}
\author[0000-0002-1264-2006]{Julianne J.\ Dalcanton}
\affil{Center for Computational Astrophysics, Flatiron Institute, 162 Fifth Ave, New York, NY 10010, USA}
\affil{Department of Astronomy, Box 351580, 
University of Washington, Seattle, WA 98195, USA}
\author[0000-0001-8416-4093]{Andrew Dolphin}
\affil{Raytheon Technologies,
1151 E. Hermans Road, Tucson, AZ 85706, USA}
\affil{Steward Observatory, University of Arizona,
933 N. Cherry Avenue, Tucson, AZ 85719}
\author[0000-0002-6301-3269]{L\'eo Girardi}
\affil{Osservatorio Astronomico di Padova – INAF, Vicolo dell’Osservatorio 5, I-35122 Padova, Italy}

\title{A Multiwavelength Classification and Study of Red Supergiant Candidates in NGC~6946}
\date{August 2020}

\newcommand{\Koplitz}{\cite{koplitz2021}}

\begin{abstract}
    We have combined resolved stellar photometry from Hubble Space Telescope (\emph{HST}), \emph{Spitzer}, and \emph{Gaia} to identify red supergiant (RSG) candidates in NGC~6946, based on their colors, proper motions, visual morphologies, and spectral energy distributions. We start with a large sample of 17,865 RSG candidates based solely on \emph{HST} near-infrared photometry. We then chose a small sample of 385 of these candidates with Spitzer matches for more detailed study. Using evolutionary models and isochrones, we isolate a space where RSGs would be found in our photometry catalogs. We then visually inspect each candidate and compare to Gaia catalogs to identify and remove foreground stars. As a result, we classify 95 potential RSGs, with 40 of these being in our highest-quality sample. We fit the photometry of the populations of stars in the regions surrounding the RSGs to infer their ages. Placing our best candidate RSG stars into three age bins between 1 and 30 Myr, we find 27.5\% of the candidates falling between 1-10 Myr, 37.5\% between 10-20 Myr, and 35\% 20-30 Myr. A comparison of our results to the models of massive star evolution shows some agreement between model luminosities and the luminosities of our candidates for each age. Three of our candidates appear significantly more consistent with binary models than single-star evolution models.
\end{abstract}

\section{Introduction} \label{intro}
Massive stars produce a large fraction of the energy and heavy elements that drive the evolution of galaxies. Red supergiant stars (RSGs) probe the late stages of this evolution, just before these stars end their lives. Some stars that form with an initial mass between 8 and 30\(M_\odot\) \citep{Massey2017} are expected to become RSGs, and many are likely to end their stellar evolution with a core-collapse supernova event (CCSNe; such as SN2017eaw in NGC~6946 \cite{Rui2019}). However, the exact supernova pathways are uncertain \citep{Coughlin2018,Smartt2015,Adams2017}

Constraining which stars produce supernovae depends critically on our ability to model the RSG stars that should produce them. As late-stage post-main-sequence stars, RSGs are shell hydrogen-burning stars with large, expanded, cool envelopes.  Their radii extend up to 1000 solar radii, leading to temperatures of 3400-4300K. Hydrogen burning in the outer layers of the star while the core is contracting causes the rapid expansion of the stellar envelope, which results in the temperature dropping. This large outer layer is made up of convective cells mixing at sonic and supersonic speeds \citep{Levesque}. 

RSG evolution is notoriously difficult to model. Rotation, binary mass transfer \citep{Raithel2018}, mass loss, convective overshooting \citep{Higgins2019}, and the compactness parameter \citep{Sukhbold2014} are all other factors that could influence which massive stars ultimately produce supernovae. \cite{Sukhbold2014} show evidence for islands of similar probability of explosion based on calculations of the compactness parameter from models. The mass ranges of stars with higher likelihood of producing supernovae (lower compactness) are 8-22\(M_\odot\) and 25-30\(M_\odot\) while stars with a mass 22-25\(M_\odot\) and above 30\(M_\odot\) show lower likelihoods.

Improving on current models of RSG evolution requires observational constraints from many massive stars. However, massive stars are rare because of their short lifetimes and the steepness of the initial mass function (IMF). By surveying an entire galaxy for populations of RSGs, we may efficiently add to the community's sample.
 
NGC~6946, the Fireworks Galaxy, is an excellent candidate for finding such evolved massive stars. It has had nine supernovae in the last century \citep{SN2013} and boasts a star formation rate between 3.2\(M_\odot\)yr$^{-1}$ \citep{Jarrett_2012} and 12.1\(M_\odot\)yr$^{-1}$ \citep{Eldridge_Xiao_2019}. One of the most active star forming and explosive galaxies should have a significant population of RSG stars to help constrain stellar evolution models.
One recent example of the usefulness of NGC~6946 for constraining massive stellar evolution was the progenitor mass measurements of \Koplitz. They present measured ages and infer masses of the progenitor of supernova remnants to constrain the mass distribution of core-collapse supernova progenitors. 

Our goal is to create a reliable sample of RSGs in NGC~6946 to improve constraints on the pre-explosion evolution of potential supernova progenitors. Through proper identification of RSGs and by estimating ages, we hope to provide critical data to supplement further work on the evolution of these massive stars. Reliably modeling RSGs and their evolution is critical to our understanding of the impact of massive stars on the evolution of galaxies. Section \ref{datas} of this paper describes the Hubble Space Telescope \emph{HST} and Spitzer data used in this project. In Section \ref{analysis} we walk through the analysis process which describes how we identify the RSG candidates and then determine their ages. Section \ref{results} lays out the results, and we discuss what we gather from the data. In Section \ref{final} we wrap up the paper with a brief summary. We assume a distance modulus of 29.4 (7.6 Mpc), consistent with the TRGB distance measurement from \cite{Anand_2018}, throughout the paper.

\section{Data} \label{datas}
In this section we describe the data and how we use it to identify RSGs. We use both \emph{HST} and Spitzer observations to search NGC~6946 for the RSGs after matching the two catalogs. We discuss what we expect model RSGs to look like, the selection criteria, and the spectral energy distributions (SEDs) used to identify the candidates. Gaia is used to look at each candidate to find which ones are foreground or mismatches between the catalogs. Finally we discuss possible low-mass contaminants such as asymptotic giant branch (AGB) stars entering our sample.

\subsection{Hubble Space Telescope} \label{data_hst}

We employed archival \emph{HST} (Version 3) imaging taken with the Advanced Camera for Survey (ACS) instrument with the Wide-Field Channel (WFC) F435W, F606W, and F814W filters along with the Wide-Field Channel 3 (WFC3) instrument with the Infrared Channel (IR) F110W and F160W filters. All the \emph{HST} data used in this paper can be found in MAST: \dataset[doi:10.17909/T97P46]{http://dx.doi.org/10.17909/T97P46}.

The HST photometry techniques used are identical to those used in \Koplitz. In short, we simultaneously measured photometry across all bands with the same point spread function (PSF) fitting pipeline used for PHAT \citep{Williams2014}, built around DOLPHOT \citep{dolphot2000, dolphot2016}. We adopted additional cuts on photometric quality to produce clean color-magnitude diagrams(CMDs) in the desired filters. These cuts are a signal-to-noise ratio greater than 4.5, sharpness squared values less than 0.4, and crowding values less than 2. The values were determined using a grid of various combinations of cuts and observing the effect on the sample. The cuts were chosen to eliminate measurements that were not a part of a stellar evolution feature on a CMD while maintaining the measurements within the features.

The HST observations are listed in Table \ref{tab:obs}. Column 1 gives the name of the image. Column 2 has the proposal number. Column 3 is the date of observation. Columns 4 and 5 have the R.A. and decl. Column 6 holds the V3 Axis in degrees. Column 7 has the exposure time. Column 8 has the filter name.

Building on \Koplitz, we included F110W and F160W WFC3/IR imaging to allow us to search for and identify RSGs. The age analysis (see Section \ref{analysis}) of the stellar population surrounding the candidate is conducted with the optical (F435W, F606W, and F814W) image data.

To allow for quantitative fitting of models to our \emph{HST} photometry, we measure the photometric completeness, bias, and uncertainties as a function of magnitude and color at each RSG position using artificial stars tests (ASTs). To do this, we again followed the same technique as in \Koplitz. These tests add a star to the imaging data with a known input brightness, color, and location, then rerun the photometry in that location. We run this routine with a sample of 50,000 ASTs covering the relevant area of color-magnitude space at each of several locations in the images. Each location was chosen to represent the quality of the photometry for any area of similar limiting magnitude and stellar density, as was done in \cite{Williams2017} and \Koplitz. Candidate locations were then grouped in photometric depth and stellar density within 0.5 mag and 0.86 stars per arcsec$^2$. These limits indicate which ASTs were applicable for each location.

\begin{figure}[!htb]
    \center{\includegraphics[width=.9\textwidth]{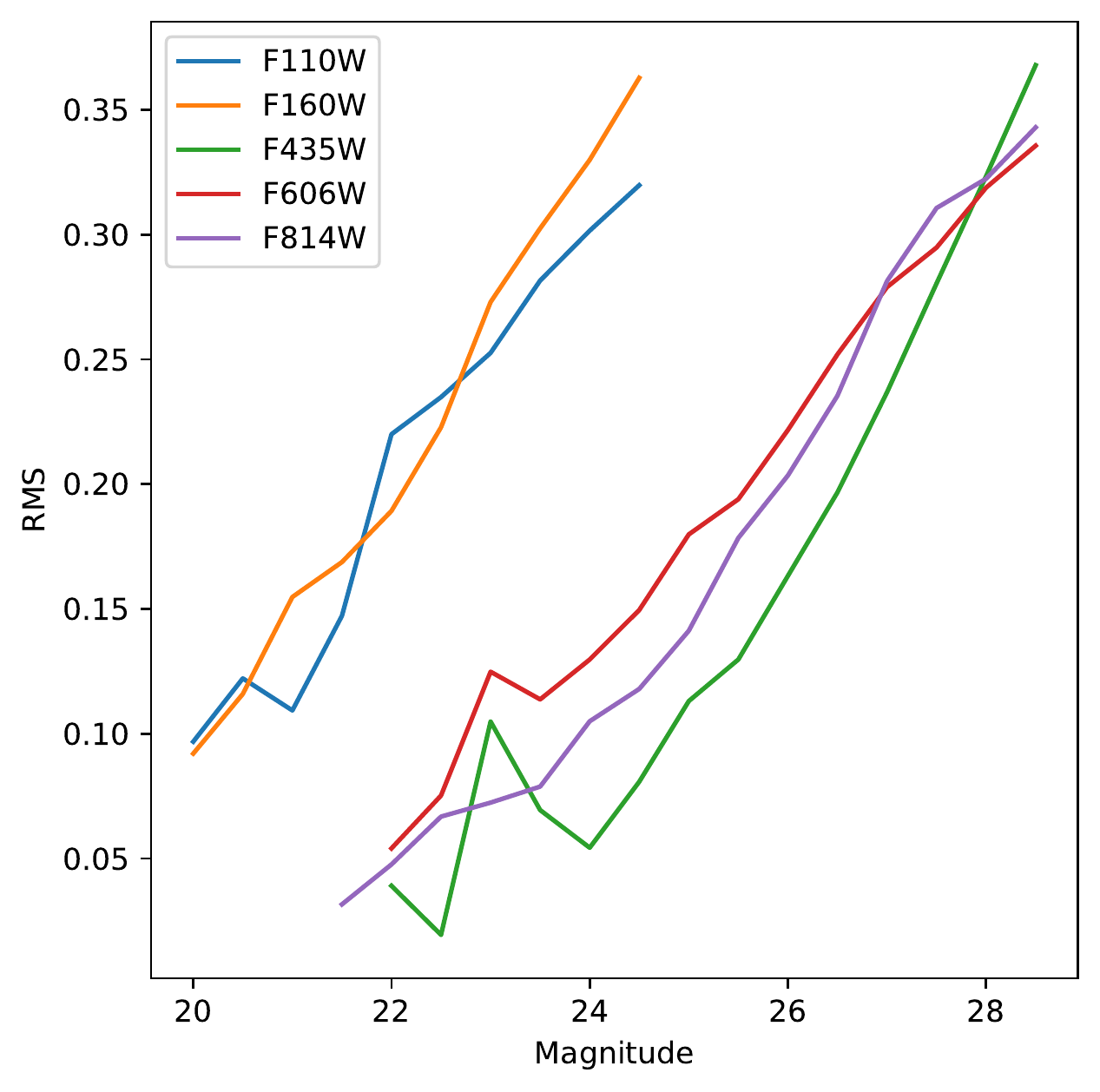}}
    \caption{\label{fig:phot_error} The RMSE of the magnitude for the ASTs. The errors in F110W and F160W, the IR filters, are similar to each other. The errors of the optical filters, F435W, F606W, and F814W, are similar to each other as well.}
 \end{figure}

The rms errors (RMSE) on the ASTs' magnitudes are shown in Figure \ref{fig:phot_error}. Errors in both the F110W and F160W filters are 0.1 at the bright end and 0.275 at the faint end.

We created an initial sample of red supergiant candidates using only the \emph{HST} data through the use of PARSEC evolutionary tracks and isochrones. We used the stellar evolution models \citep{Bressan2012,Bressan2013,Chen2014,Tang2014} and computed isochrones using the collaborative CMD 3.3 input form (http://stev.oapd.inaf.it/cgi-bin/cmd, \cite{Bressan2012,Bressan2013,Chen2014,Chen2015,Tang2014,Marigo2017,Pastorelli2019}). These references are for all the information that went into the isochrone calculation, all of which can also be found on the CMD input form. The only settings changed were the age range, which we set to .1-40 Myr (in 100,000 yr increments), and the filter set. This form creates isochrones with magnitudes in a specified filter set, so we created isochrones for both the \emph{HST}/WFC3 and Spitzer filter sets. To correct for reddening, we add the distance modulus of 29.4 to the magnitudes in every filter, and then we add the respective galactic extinction of the filter \citep{Schlafly2011}.

\begin{figure}[!htb]
    \center{\includegraphics[width=\textwidth]{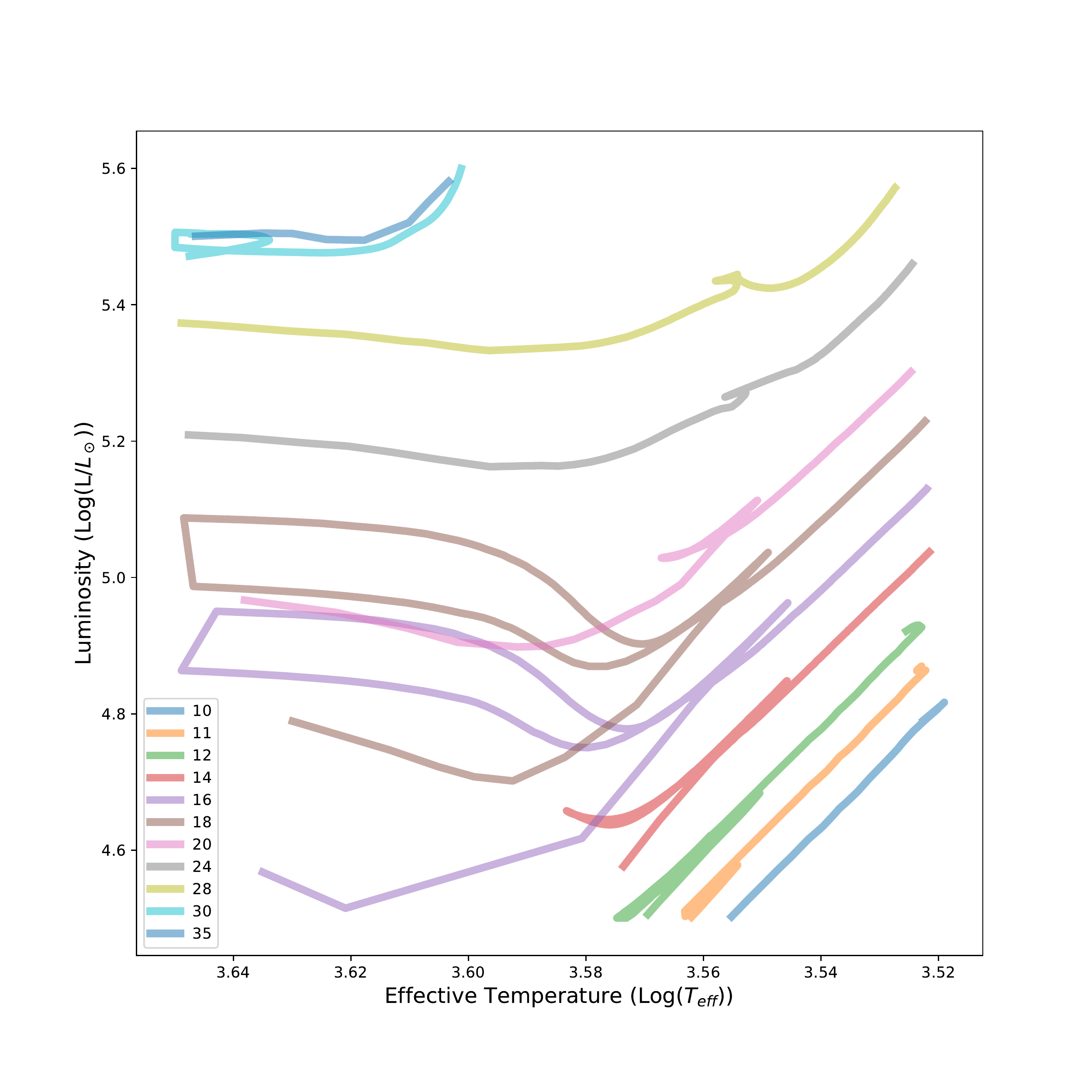}}
    \caption{\label{fig:evo_track} The evolutionary tracks (PARSEC) of stars with a given mass. We used effective temperature limits of $3.5<T_{eff}<3.65$ and luminosity limits of $4.5<Log(Lum)<5.6$. The age range of the stars that pass these cuts is 5.2-26.2 Myr. This area shown in the plot represents a space of effective temperature and luminosity space that will likely select for RSGs.}
 \end{figure}

Model RSGs have effective temperatures of 3400-4300K, or Log($T_{eff}$) of 3.53-3.63 \citep{Levesque}. Figure \ref{fig:evo_track} shows evolutionary tracks of ~10-35\(M_\odot\) stars entering this temperature range with luminosities of $4.5<Log(L_\odot)<5.6$. AGB and Super-AGB stars have initial masses of 8-9\(M_\odot\) so we avoid contamination from these stars by isolating tracks of stars with an initial mass $\geq$ 10\(M_\odot\). Applying the effective temperature, luminosity, and mass limits to the evolutionary models shows us that stars between the ages of 5.2 Myr and 26.3 Myr enter the space defined by our limits. These limits also give us a cap on the initial masses of these stars, making the mass range 10-35 \(M_\odot\) (see Figure \ref{fig:evo_track}).

\begin{figure}[!htb]
    \center{\includegraphics[width=\textwidth]{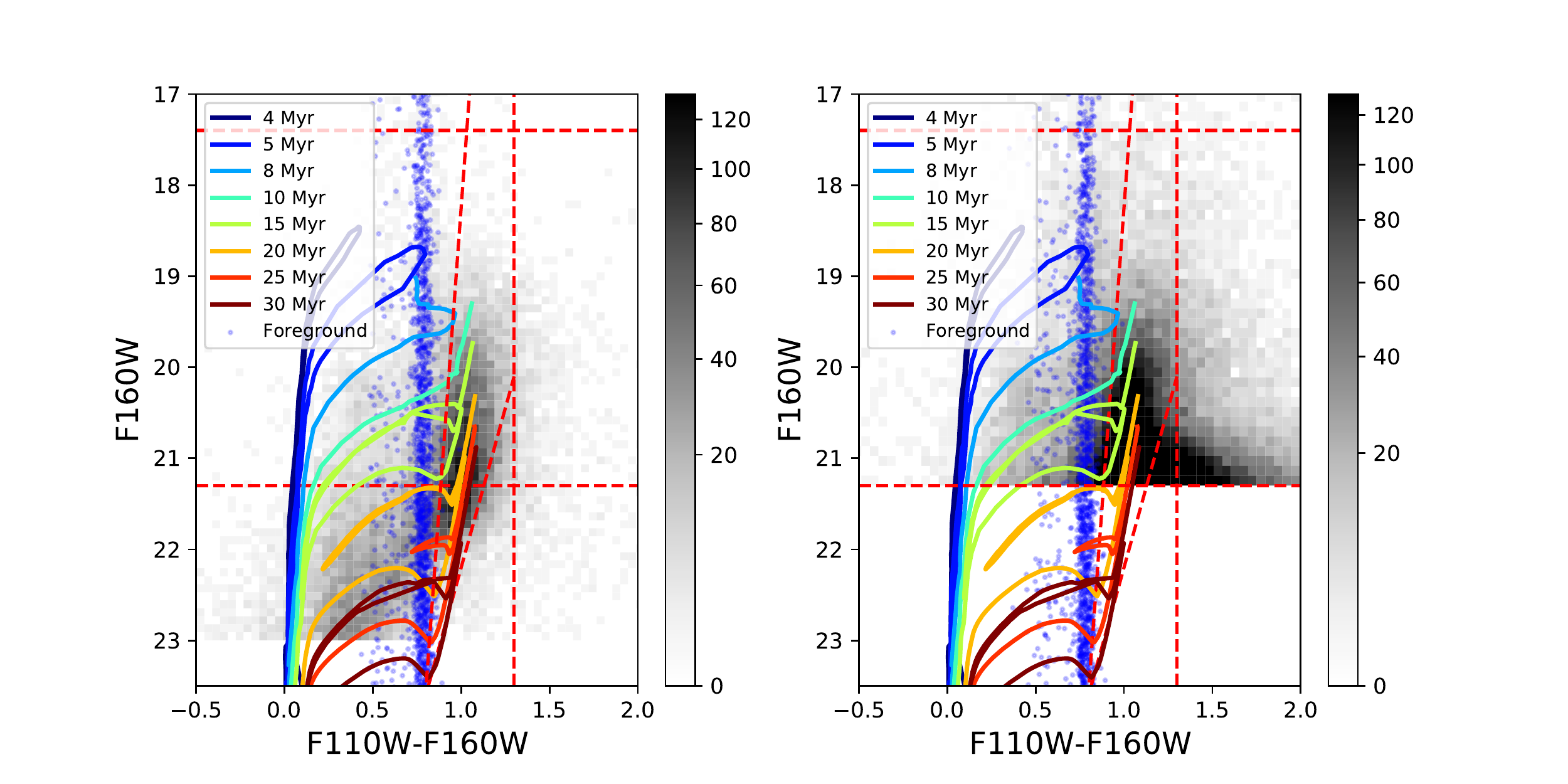}}
    \caption{\label{fig:earlycmd} CMDs of the \emph{HST} photometry data catalogs. The y-axis is the magnitude of the sources in F160W and the x-axis is the F110W-F160W color of each source. The plot on the left has the histogram CMD of our reduced \emph{HST} catalog after removing sources fainter than the 24th mag in F814W and the 23rd mag in F160W. We did not have complete coverage in F814W but we used these cuts to identify the feature toward the tip of the isochrones. This helped us shape the cuts we used on the overall catalog to find the RSG candidates. The plot on the right shows all the sources detected in the IR filters of our entire original catalog. A rainbow of differently aged isochrones (with solar metallicity) is plotted over the CMDs. The blue points are the foreground stars generated from the TRILEGAL model. We shape our cuts to avoid the majority of foreground contaminants.} Most notably, we see a high density of \emph{HST} sources around the red tip of the 10-30 Myr isochrones. The cuts we end up making to isolate that space around the isochrone tips are shown as red dotted lines.
 \end{figure}

We then take the age range determined from the evolutionary tracks and employ the isochrones for stars in that range. Figure \ref{fig:earlycmd} shows IR CMDs of F110W-F160W vs F160W as two-dimensional histograms. The CMD points represent our data after the processing and filtering steps described above. Layered on top of the CMD are the isochrones of different ages along with model foreground stars from TRILEGAL \citep{Vanhollebeke2009, Girdari2007}.

The input parameters for the TRILEGAL models follow the formulas in Table 1 of \cite{Pieres2020} and are laid out below:
\begin{itemize}
  \item Dust: Av$_\infty = 1.04 \pm 0.06$ mag (from the Schlegel maps, \cite{Schlegel1998}), scale height = 110 pc
  \item Thin disk: exactly the same parameters as in the table
  \item Thick disk: surface density = 0.0010 $M_{Sun} / pc^2$, scale length = 2913 pc, scale height = 800 pc (these were values previous to the recalibration done in \cite{Pieres2020})
  \item Halo: a single power-law (and not two power laws as in \cite{Pieres2020}) with local density = 0.0001 $M_{Sun} / pc^2$, effective radius = 2698.93 pc, oblateness = 0.62, exponent = 2.75.
\end{itemize}

These values are the result of a recalibration with SDSS and 2MASS photometry mentioned in \cite{Girardi2012}; however, the recalibration is a minor change with respect to the original calibration well documented in the original paper at \cite{Girardi2005}.

The isochrones of ~8-30 Myr stars show RSGs around an F110W-F160W color of 1. The isochrones and TRILEGAL foreground models allow us to make the following criteria that we can apply to our sample to avoid most of the foreground stars and eliminate unlikely RSGs; the cuts are shown in Figure \ref{fig:earlycmd}:

\begin{enumerate}
    \item 17.4 $<$ F160W $<$ 21.8
    \item -26.5$\cdot$(F110W-F160W)$ + $44.75$
    < $F160W$ < -7\cdot$(F110W-F160W)$ + $29.2
    \item F110W - F160W $<$ 1.3
\end{enumerate}

\begin{figure}[!htb]
    \center{\includegraphics[width=\textwidth]{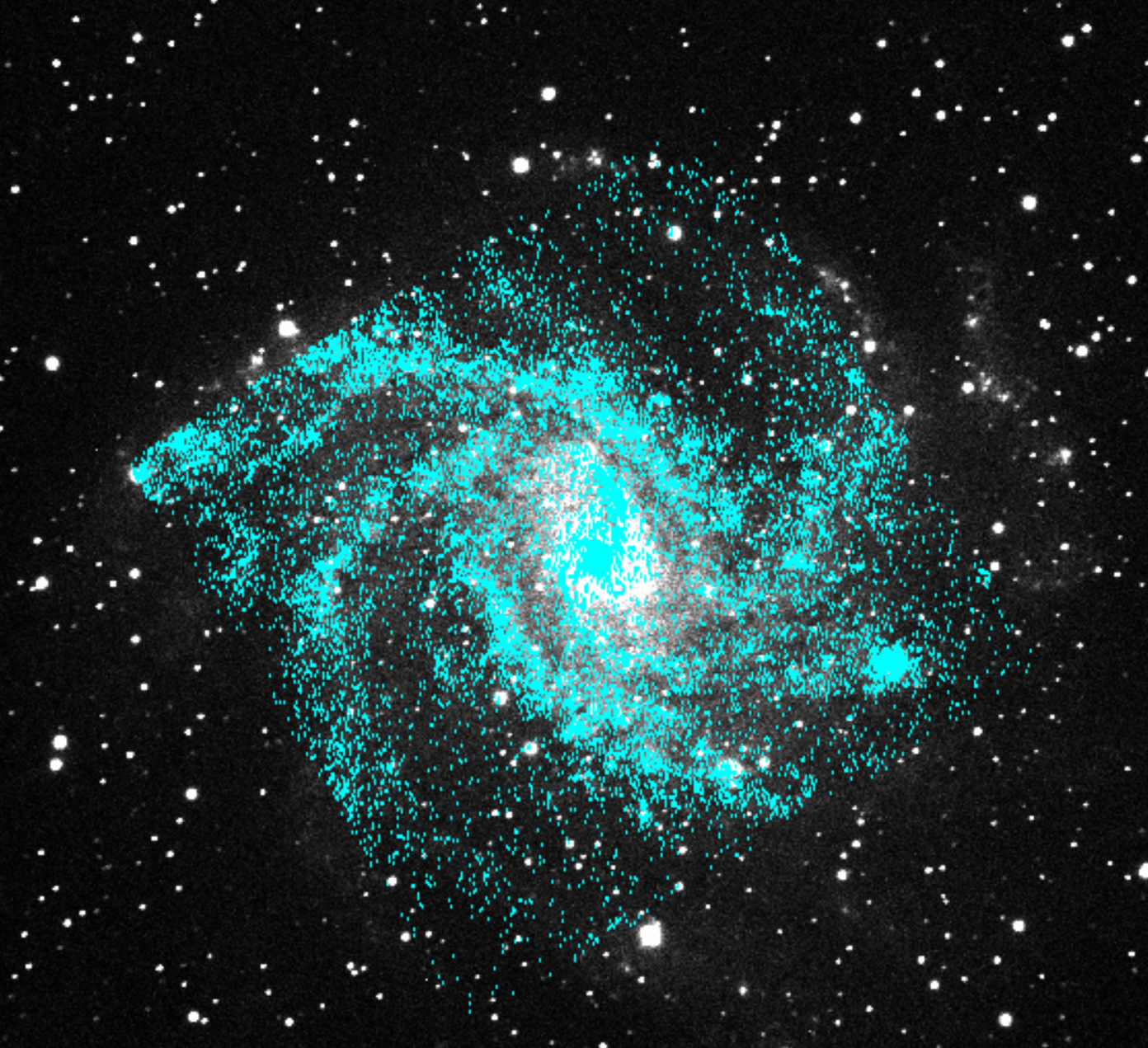}}
    \caption{\label{fig:IRgalaxy} A KPNO2.1m image of NGC~6946 \citep{Kennicutt2003} overlaid with the locations of the '\emph{HST}-only' sample. This figure shows how these stars are grouped around the spiral arms of the galaxy as you would expect for RSGs.
     North is up and east is to the left.}
 \end{figure}

This produced a sample of 17,865 stars which we will refer to as the '\emph{HST}-only' sample. The positions and magnitudes of these stars are provided in Table \ref{tab:IRPhot}. Figure \ref{fig:IRgalaxy} shows our large \emph{HST}-only sample mapped onto an image of NGC~6946. The positions of these stars line up very well with the spiral arms of the galaxy, which is expected when looking for young RSGs. To focus our sample and find an optimal set of the top candidates for follow-up study, we use additional information from Spitzer.

\subsection{Spitzer Photometry} \label{match}

We supplemented our \emph{HST} photometry with Spitzer bands using  the catalog from \cite{Khan_2017}. The photometry was taken with the 3.6, 4.5, 5.8, 8 $\mu$m filters on the Infrared Array Camera (IRAC; \cite{Fazio2004}) on board the Spitzer Space Telescope \citep{Werner2004}. RSGs are bright in the Spitzer bands such as 3.6$\mu$m and separate from other stars in some regions of color space \citep{BonanosAZ2009}.

We cross-correlated the Spitzer and \emph{HST} photometry catalogs to ensure the same source was being detected in each catalog at the same location. Red supergiants are typically bright at mid-infrared wavelengths \citep{Messineo2012}. We therefore used the available Spitzer point-source catalog for NGC~6946 to find stars in our \emph{HST} catalog with bright mid-infrared counterparts. Because the Spitzer resolution is much coarser than the resolution of \emph{HST}, this matching was not straightforward. The original Spitzer catalog contained 15,815 sources but covered an area larger than our \emph{HST} catalog did. After eliminating sources outside of our \emph{HST} coverage area we have 7918 total Spitzer sources. We expect the majority of these to be AGB or RGB stars, which are far more numerous than more massive RSGs or background galaxies. 

\begin{figure}[!htb]
    \includegraphics[width=.5\textwidth]{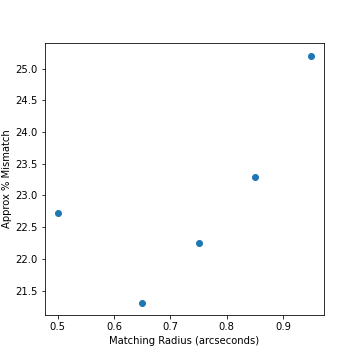}
    \includegraphics[width=.5\textwidth]{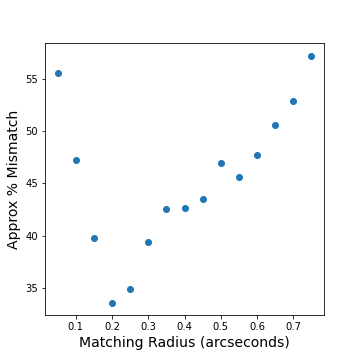}
    \caption{\label{fig:radius_graph} The approximate percentage of mismatches at each radius. The approximate number of mismatches is shown on the y-axis and is computed by taking the number of matches when the RA coordinates were shifted then dividing the number of matches when using the actual coordinates. The radius used is shown on the x-axis. The left plot shows the matching radius used to create the primary sample with 0".65 (13 pixels) as the most optimal radius. The plot on the right shows the best matching radius of 0".2 (4 pixels) for the secondary sample of RSG candidates.}
 \end{figure}

We estimated the best matching radius by identifying the radius with the lowest fraction of spurious matches to real matches. To address the simplest case, we started by limiting the cross-correlation to one-to-one matches. We tested radii ranging from 2 to 100 pixels on our \emph{HST} reference image, where 1 pixel is 0."05. We evaluated the rate of spurious matches for each radius by shifting the Spitzer coordinates by 8" and rerunning the matching procedure, such that all 'matches' in the shifted case should be spurious. Figure \ref{fig:radius_graph} shows the fraction of spurious sources as a function of matching radius. There were differences in matching success depending on the density of surrounding sources. We determined an optimal matching radius of 13 pixels (corresponding to 0".65 at 0".05pixel$^-1$) for candidates with only one HST source and one Spitzer source within the radius. This resulted in a 21.5\% mismatch rate. There were no locations with multiple Spitzer sources within 0".65. 385 paired HST and Spitzer sources were found at this 0".65 radius and became our primary sample of RSG candidates.

While 0".65 was found to be optimal for one-to-one matches between catalogs, in more crowded regions of the HST images, a 0".65 search radius matched more than one HST source to some Spitzer sources. Therefore, we determined a separate limiting radius for sources with more than one match within the initial 0".65 search.  We optimized this radius by finding the lowest fraction of spurious to real matches within only these crowded regions. For candidates with multiple \emph{HST} sources within 0".65 of a Spitzer source, a smaller radius of 4 pixels (0".2) was optimal for cross-correlation in a crowded space. If there were still multiple \emph{HST} sources within this 0".2 radius around a Spitzer source, the closest \emph{HST} source was matched. Additionally, if there were multiple \emph{HST} sources within 0".65 but none within 0".2, the closest \emph{HST} source was matched to the Spitzer source. The technique of using a 0".2 matching radius and taking the closest match in multiple-match cases had a higher mismatch rate of about 30\%. The secondary radius technique resulted in 124 paired sources.

To obtain the highest quality sample of RSG candidates, we focus on our primary 'single' match sample of 385. The secondary sample of 124 candidates is included in the tables and a figure, however we keep our analysis focused on the former sample. Both samples are put through selection criteria laid out in the rest of this section. The rmse on the Spitzer source positions is 0".327 R.A. and 0".242 decl.
  
\subsection{Multiwavelength Culling of the Sample} \label{identify}

We make additional cuts using Spitzer filters to narrow down to the most likely RSG candidates, based on existing studies of the SEDs of RSGs. We use a mix of selection criteria from \cite{Britavskiy2014}, including the F110W - [3.6] $>$ 1 cut, which they also use to select for RSGs, and from the isochrones as described above. The criteria in \cite{Britavskiy2014}; based on work from \cite{BonanosAZ2009}) used the J filter but we substituted the overlapping F110W filter.

The first three criteria below are mentioned above while the latter two are from \cite{Britavskiy2014}.

\begin{enumerate}
    \item 17.4 $<$ F160W $<$ 21.8
    \item -26.5$\cdot$(F110W-F160W)$ + $44.75$
    < $F160W$ < -7\cdot$(F110W-F160W)$ + $29.2
    \item F110W - F160W $<$ 1.3
    \item F110W - [3.6] $>$ 1
    \item $[3.6] - [4.5] <$ 0
\end{enumerate}

After combining all criteria and applying them to the 385 possible candidates, 24.4\% of the candidates remain, which leaves us with 94 RSG candidates in the primary sample. We expect roughly 20 mismatches. A potential explanation for so few candidates passing these cuts is that we initially found many background galaxies. These galaxies will not have the colors we are selecting for and have now been removed from the sample. These same cuts were also applied to the secondary sample of 124 candidates and reduced the sample to 43 candidates.

\begin{figure}[!htb]
    \center{\includegraphics[width=\textwidth]{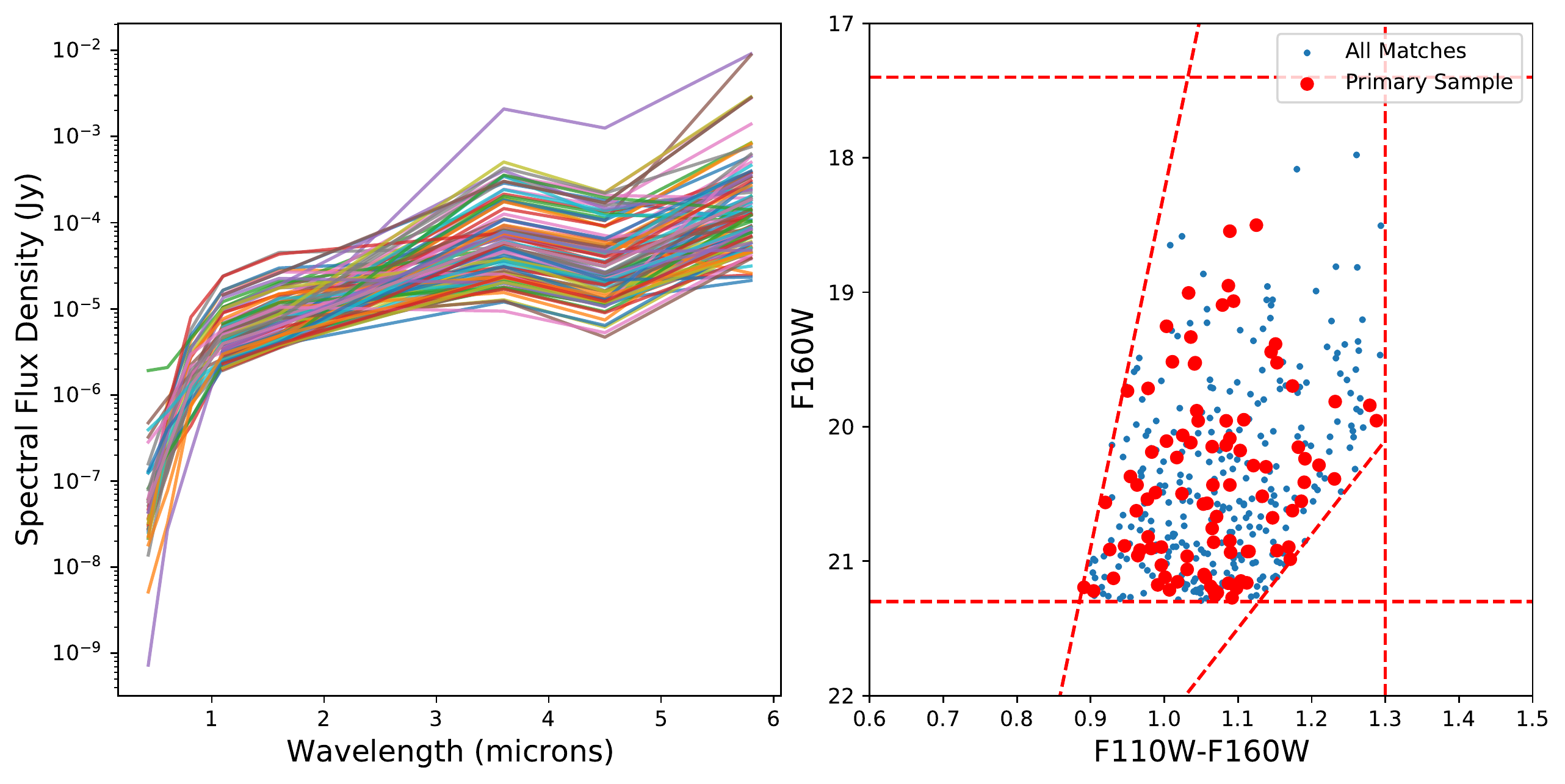}}
    \caption{\label{fig:SED} The left panel shows the SED of the stars that pass all of our photometry cuts on both \emph{HST} and Spitzer, our primary sample. The y-axis is the spectral flux density in Janksys. The x-axis is the wavelength of light. The right panel is a CMD showing some of the cuts we used as red dotted lines. The y-axis is the magnitude of the stars in F160W and the x-axis is the F110W-F160W color of each star. We see the majority of our primary sample stars with similar SEDs. The candidates with a distinct peak at 3.6$\mu$m are blends of multiple stars, which are likely mismatches or foreground stars.}
 \end{figure}

We plot the candidates' SEDs using our optical-mid-IR photometry in Figure \ref{fig:SED}. This is used to check for relative uniformity between the SEDs of our candidates. We converted \emph{HST} and Spitzer magnitudes into flux values using the zero magnitude flux values ($F_0$) for \emph{HST} from the SVO Filter Profile Service (\cite{Rodrigo2012,Rodrigo2020}; 4197.72, 3250.63, 2421.40, 1702.00, and 1132.88 for the F435W, F606W, F814W, F110W, and F160W filters, respectively) and used the IRAC Instrument Handbook \citep{IRAC2.2, IRAC4.3} to get the values for Spitzer (280.9, 179.7, and 115 for the 3.6, 4.5, 5.6 micron filters, respectively).

\begin{figure}[!htb]
    \center{\includegraphics[width=\textwidth]{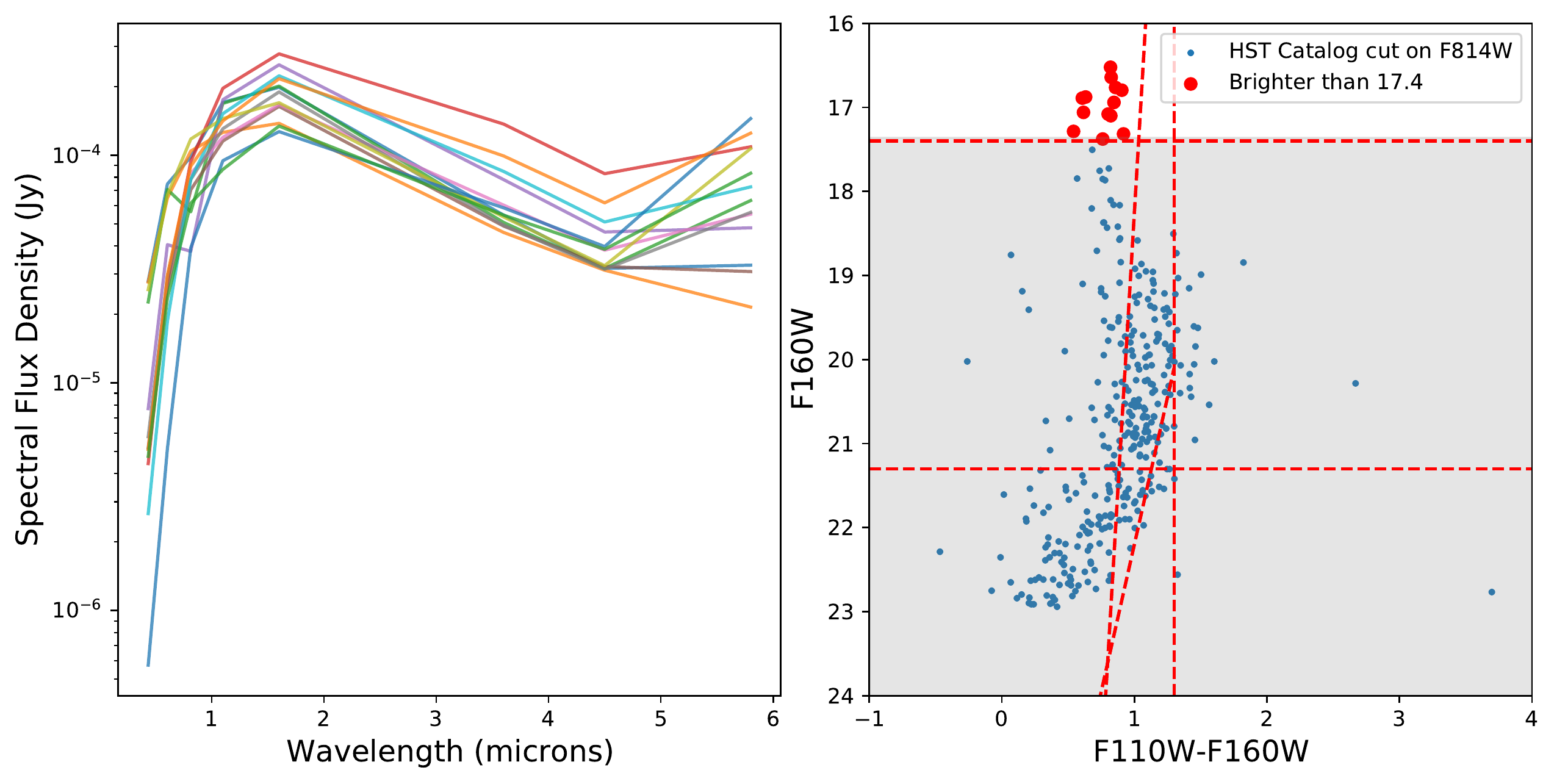}}
    \caption{\label{fig:SED_f} Similar to Figure \ref{fig:SED} except the F160W cut at a magnitude of 17.4 in F160W is reversed and picks the stars above the red dotted line. This shows that these stars have a different SED shape than in Figure \ref{fig:SED}.}
 \end{figure}

The SEDs in Figure \ref{fig:SED} show excess flux toward the mid-IR associated with dust-enshrouded RSG stars. There is an example in Figure \ref{fig:SED_f} showing SEDs of stars brighter than F160W magnitude 17.4. These stars show a peak flux density of 1.6$\mu$m, a much different shape than our primary sample SEDs. The criteria adopted to identify RSGs are not perfect, so some stars shown in the SED could be foreground stars and some stars eliminated from the sample could be RSGs.

A compilation of all the magnitudes for each RSG candidate is shown in Table \ref{tab:Phot}. The first column is a unique identifier for each RSG candidate. The identifier is the index of the star in the Spitzer data. Columns 2, 3, 4 and 5 give the R.A. and decl. for \emph{HST} and Spitzer. Columns 6, 7, 8, 9, and 10 have the magnitudes of the \emph{HST} filters F435W, F606W, F814W, F110W, and F160W respectively. Columns 11, 12, and 13 have the magnitudes of the Spitzer 3.6$\mu$m, 4.5$\mu$m, and 5.8$\mu$m filters. Below we discuss other potential contaminants to this sample, which results in many of them being flagged as less likely candidates in Table \ref{tab:MATCH_results} (described in the introduction to Section \ref{results}).

Pairing \emph{HST} and Spitzer sources resulted in a substantial decrease in the size of our viable sample. The much higher resolution of \emph{HST} helps produce a clean sample of individual stars while Spitzer helps to identify bright red stars in its NIR filters. Because of the additional information from Spitzer, we are most confident in labeling the primary and secondary samples as our possible RSG candidates.

We calculated the approximate fraction of RSGs in NGC~6946 represented in our sample by estimating how many red supergiants we expect to see in the galaxy based on the star formation rate. Using a star formation rate of 3.2\(M_\odot\)yr$^{-1}$ \citep{Jarrett_2012} over a period of 30 Myr, we expect roughly 96 million solar masses to be formed. Assuming a Kroupa IMF and integrating between our mass range of 10-35\(M_\odot\) we estimate 400,000 RSGs forming in NGC6946 over a 30 Myr period. From the PARSEC evolutionary tracks the lifetime of RSGs is $\sim$600,000 yr, or about 2\% of these 30 Myr. Thus, we may expect $\sim$8000 RSGs to be present in the whole galaxy. Therefore, our \emph{HST}-only sample likely has about 50\% contamination; however, our small primary sample, which has very few stars with significantly more vetting, is likely to be much cleaner. The size of our primary sample is not surprising given the high degree of crowding of the Spitzer data. High-resolution mid-IR photometry, perhaps from future observations with the James Webb Space Telescope, will likely yield many more high-quality RSG candidates.

\subsection{Contamination}

We further culled our sample of RSG candidates by visual inspection and comparison with Gaia. Using the Gaia Single Source service \citep{Gaia2016, Gaia2018}, we were able to visually inspect each location and compare the R.A. and decl. of the \emph{HST} and Spitzer detections, and determine likely mismatches as well as foreground stars. Each candidate had different coordinates for the \emph{HST} and Spitzer detections. With these visual inspections, we found 19 locations where a much brighter nearby star would more likely be the Spitzer detection than the fainter \emph{HST} source from the sample. This was slightly below the number of mismatches we expected. These candidates were labeled as mismatches between the \emph{HST} and Spitzer catalogs. Furthermore, another 21 candidates had measured proper motions in Gaia, and therefore are likely foreground stars. Removing these likely mismatches and foreground stars brings us to 54 RSG candidates. The removed candidates are labeled in Table \ref{tab:MATCH_results} with the "M" and "F" flags for mismatches and foreground. The secondary sample was reduced to 11 candidates after removing the mismatches and foreground stars. We found 22 mismatches, and 10 foreground stars were identified in the secondary sample. We expected 13 mismatches from this sample. 

\begin{figure}[!htb]
    \center{\includegraphics[width=\textwidth]{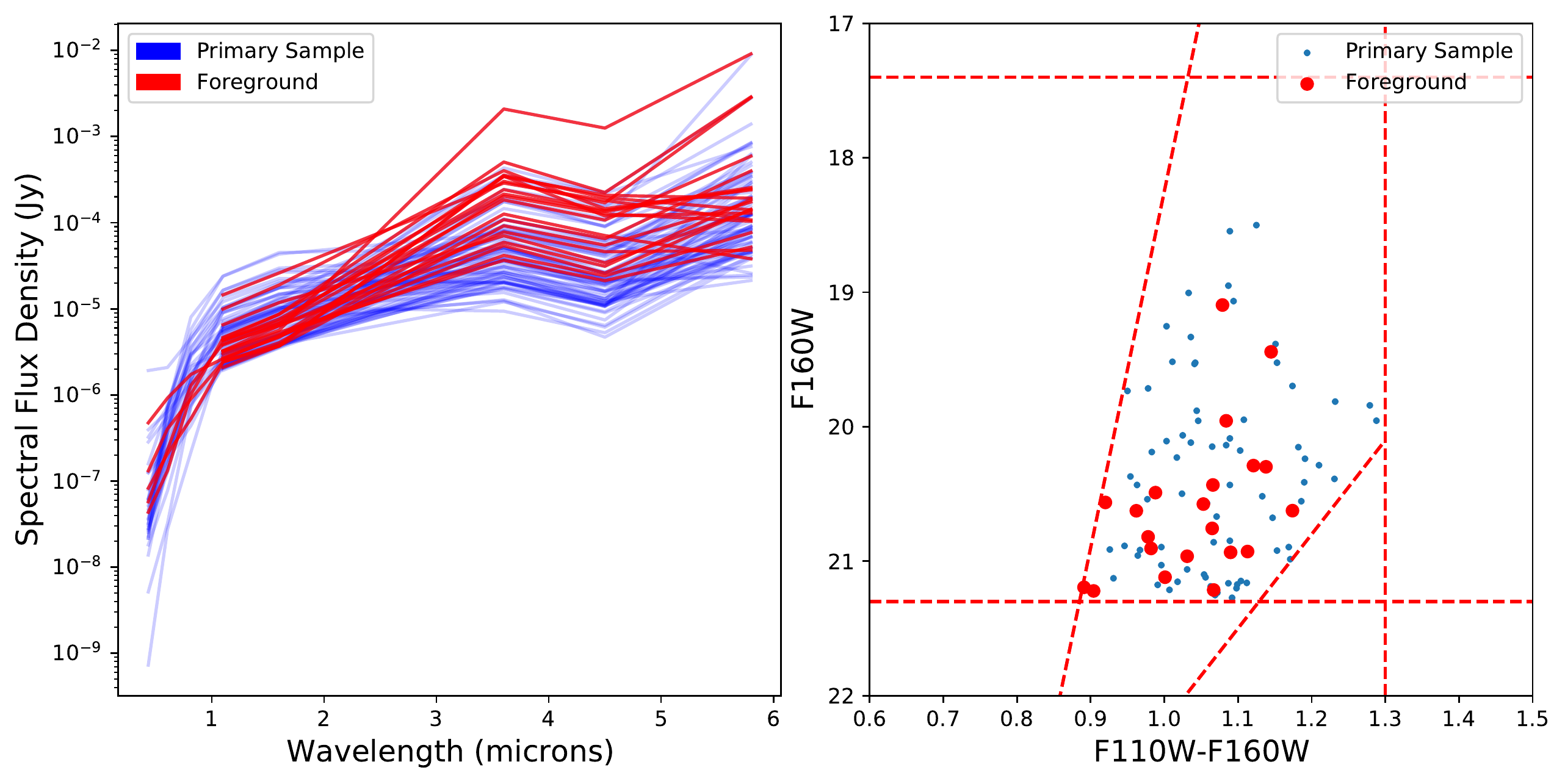}}
    \caption{\label{fig:SED_m} This SED shows our sample (blue) and identified foreground stars (red). The foreground SEDs are not easily separated from our primary sample SEDs, but were identified using other methods. Most of the stars with the peak at 3.6$\mu$m were identified as foreground stars.}
 \end{figure}

Identifying some very likely foreground stars in our sample allowed us to plot their SEDs to compare against the rest of the sample. Figure \ref{fig:SED_m} shows SEDs color coded according to the group they belong to. Our primary sample is in blue and the identified foreground stars in red. We found that the blending of multiple stars could be responsible for SEDs peaking at 3.6$\mu$m. The vast majority of these SEDs belonged to mismatched or foreground stars. The Spitzer resolution is much lower than HST, so multiple resolved stars in the HST catalog may be only a single, erroneously bright, Spitzer detection.
  

In addition to contamination from the foreground, there could be bright red stars in NGC~6946 that are not RSGs, such as AGB stars. Although our luminosity selection is brightward of the evolutionary pathways of nearly all low-mass stars, AGB stars entering our defined temperature and luminosity space is still a major concern. The possible AGB contaminants are labeled as such with the "C" flag in Table \ref{tab:MATCH_results} identified by their cooler temperatures and lower luminosities. The candidate locations now labeled possible AGB contaminants all returned star formation within our accepted age ranges making them still viable RSG candidates as well. A figure showing these possible contaminants compared to our selection criteria and results is discussed later in Section \ref{p&g_models}.

\section{Analysis} \label{analysis}
The overall goals of our analysis were to find a reliable set of RSG candidates in NGC~6946, constrain their ages, and compare their ages and photometric properties to stellar evolution models. In the following subsections, we will first discuss using MATCH to determine ages, and then the existing models we will use to compare against our results. 

\subsection{Constraining RSG Candidate Ages} \label{use_match}
Previous work has shown that some of the young stars within a 50 pc radius from a point are likely to be coeval, allowing the neighboring stars to be used for age dating \citep{Gogarten2009, Williams2014hist}. This is especially true for younger stars that have less time to travel away from their original population. From \cite{Bastian2006}, stars formed from a common event maintain a spatial relationship at distances up to 100pc for about 100 Myr. Applying the initial velocity dispersion, usually a few km$^{-s}$, the stars bound to escape the cluster are still associated with the cluster for 15-45 Myr. It is therefore likely that most young stars in a 50 pc area have coevolved, although not all stars. We take multiple steps to reduce possible contamination within our targeted location including foreground stars as described above and the use of background CMDs, which we discuss later in this section. Therefore, we can infer the likely ages of the RSGs by analyzing the surrounding stellar population. To determine an age distribution for each of the RSG candidates we used MATCH \citep{Dolphin2002, Dolphin2013}. This program models the CMDs to infer star formation history (SFH) for that stellar population. The resulting SFH from MATCH includes a distribution of ages, metallicities, and uncertainties on these values. We use the distribution of ages as a proxy for the probability that the RSG candidate is of a certain age.

MATCH applies two types of extinction to the model CMDs when fitting the data. These are called $dA_{v}$ which represents the differential extinction within the region being measured and $A_{v}$ which accounts for the uniform foreground extinction in the region. These $dA_{v}$ and $A_{v}$ values were solved for by inputting a range of values and MATCH would choose the most likely value corresponding to the best fit. We tested fits to the data with a grid of possible $dA_{v}$, and $A_{v}$  to determine the best combination for each region. To determine the $A_{V}$ value we calculated the best fit to the observed CMD, testing values greater than 0.8 (slightly below the foreground value of \cite{Schlegel1998}) in 0.05 increments.
We also supplied MATCH with some restrictions on metallicity to limit the  parameter space of the fitting. \cite{Cedres2012} found an NGC~6946 metallicity that is approximately solar. A range of metallicities from $-0.5 \leq$ [Fe/H] $\leq 0.1$ was used as the MATCH input. The most consistent metallicity results from MATCH are [Fe/H] = $0.05$, which is also consistent with the metallicity of the models used in this paper, described later in Section \ref{models}.
Finally, we applied background CMDs to isolate the unique population associated with the positions of the RSG candidates. These background CMDs are taken from an annulus extending from 50 pc to 1 kpc around each source. We then weighed the contamination CMDs to the area of the original radius. This contamination CMD also helps to correct for foreground issues as these stars are also evenly distributed across the face of the galaxy. Moreover, the main sequence of young stars in NGC6946 is bluer than the foreground stellar sequence.

Following the technique used by previous work in determining ages using MATCH age distributions \citep{Jennings2012,Williams2019}, we find the median age of the young stars in the best fit for the sample at each location. We then assign that as the most likely age of the RSG candidate. We restrict the maximum age to 30 Myr because the models (described in Section \ref{p&g_model_info}) show that the maximum age of RSGs is $<$30 Myr. Thus, older ages present are not likely to have stars remaining that are massive enough to become RSGs. If the location around an RSG candidate does not show significant star formation at an age 30 Myr or younger, we assign an age of zero. The uncertainty on the age is calculated by resampling the age distribution output, applying draws from the measured uncertainties, recalculating the best-fit age, and analyzing the distribution of resulting best-fit ages. More details on this technique of estimating age uncertainties can be found in \cite{Williams2018,Williams2019}. We tested the effect of offsetting the position of the RSG candidate to see how that would change the number of stars in the neighborhood of the candidate. We tested an offset of ~0".25 and found that the sample of stars around each candidate typically changed by $<$ 4\%. We reran these offset locations through MATCH and found no change in the best-fit ages.

To create the input CMD for MATCH, all \emph{HST} sources in a 50 pc radius around each candidate location were put into separate catalogs, excluding the actual RSG candidate. Because the infrared has poor sensitivity to the upper main sequence stars, which are blue, and because crowding limits the depth of the IR imaging, we were only able to constrain the population ages for regions that had complimentary optical \emph{HST} imaging. Since there is only optical coverage of 2\% of our \emph{HST}-only sample, this subset could produce some bias in the age analysis.  However, because the optical coverage is not focused on any particular galaxy feature (e.g., the center or a single star-forming region), such a bias should be minimal.

\begin{figure}[!htb]
    \center{\includegraphics[width=\textwidth]{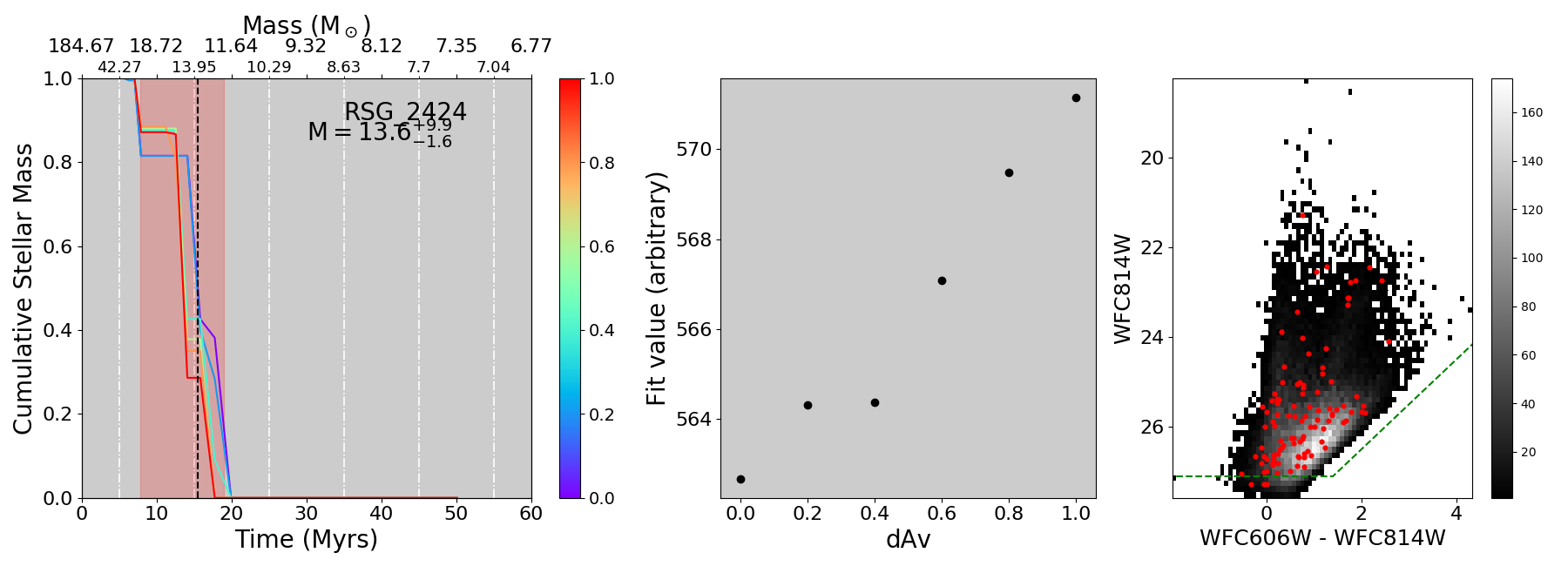}}
    \center{\includegraphics[width=\textwidth]{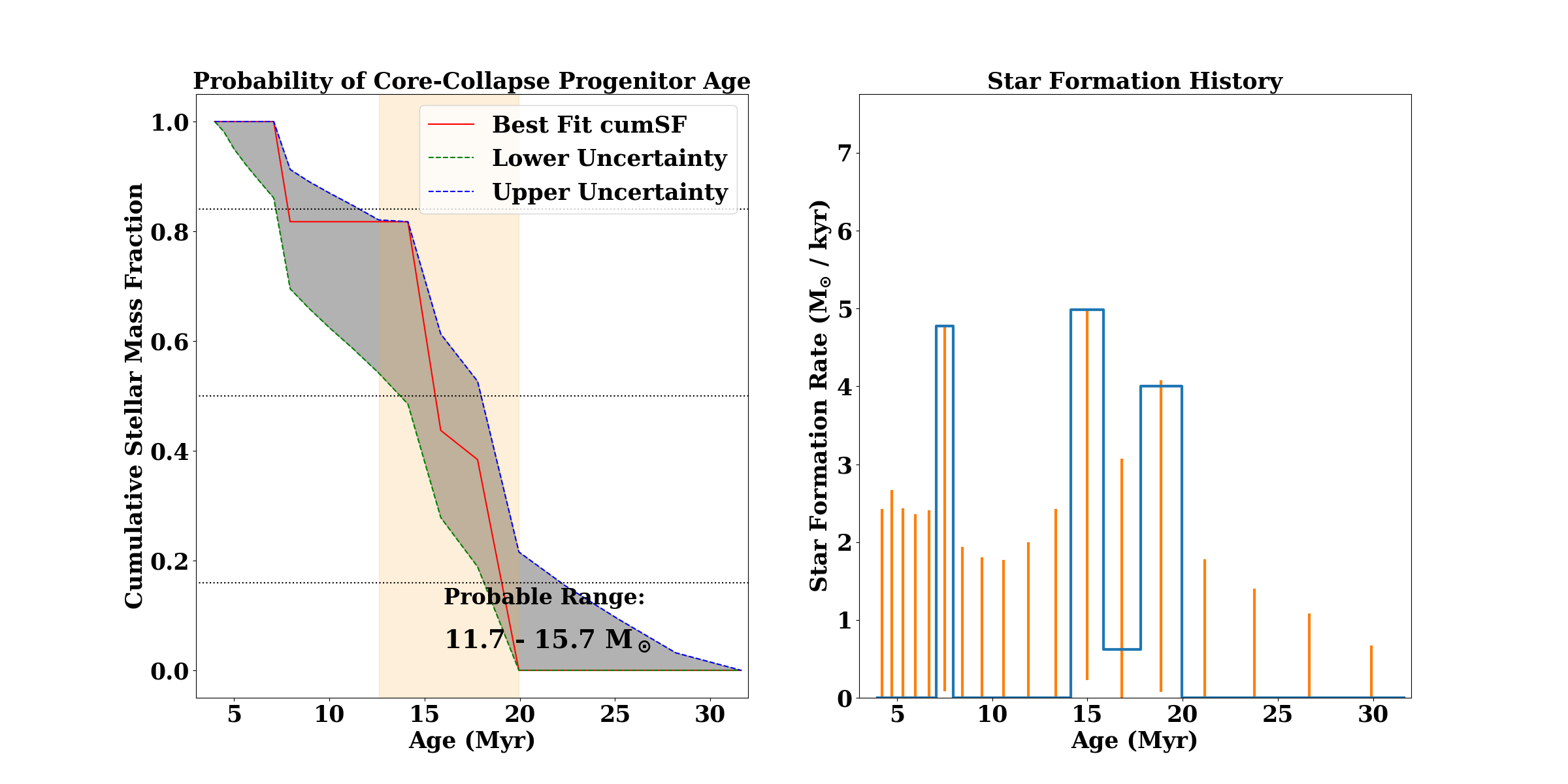}}
    \caption{\label{fig:MATCH_graphs} The top three plots are examples of the output figures from MATCH. This example is from the results of candidate RSG2424. The plot on the left shows the cumulative stellar mass at different ages of the population with a most likely age shown as the dotted black line. The rainbow lines show the various fits that MATCH tried using different $dA_{v}$ values. MATCH also gives a mass estimate (top x-axis) in this plot that we do not pay attention to for the purposes of this paper. The middle plot shows the best $dA_{v}$ value. The plot on the right is the observed CMD from the extraction photometry. The y-axis is the magnitude of the stars and background sources in the redder F814W filter. The x-axis is the color of these stars and sources in the F606W-F1814W combination. The red points are the magnitudes of the actual stars in the location that} we fed into MATCH to determine SFH. The green dashed line indicates the completeness limits determined by MATCH during the fitting. The gray-scale points are the background sources used to determine the expected population. The bottom set of plots represents a rerun of the output from MATCH in order to determine better constraints on the uncertainties. The method of determining uncertainties is described in \cite{Williams2018}. The left plot is similar to the left plot on the top but focuses on the best fit, the uncertainties on that fit only, and the age estimate from the fit. The plot on the right shows the star formation rate at different ages from the population and the uncertainties. 
 \end{figure}

There are parts of NGC~6946 that we do not have F814W coverage for. We analyzed 15 stars from the primary sample using the F435W and F606W filter combination. The remaining 32 stars from the primary sample and the entire secondary sample were analyzed using the F606W and F814W combination. We found seven locations from the primary sample, that were not already removed, which had \emph{HST} optical coverage in only one band. This coverage was insufficient to be analyzed using MATCH. Thus, although we did not have optical coverage over the entire galaxy, the areas with insufficient data only resulted in seven fewer age measurements. MATCH outputs an SFH for each $dA_{v}$ value attempted. We calculate the most likely age using the SFH from the best fit $dA_{v}$. In Figure \ref{fig:MATCH_graphs} we show an example of one CMD fitted by MATCH, its output SFR, and uncertainties on that SFR. The figure shows the cumulative stellar mass with a most likely age, the best $dA_{v}$ value, and the CMD fitted by MATCH with both real magnitudes and background fake magnitudes. The figure also contains a rerun of the MATCH output to better constrain the uncertainties on the SFH. MATCH takes into account both the photometric uncertainties when fitting the models.  In short, MATCH determines the photometric uncertainties and completeness for each color and magnitude bin from the AST results, and it applies these measurements to the models during the fitting.  Then, during the Monte Carlo error analysis, both the uncertainties and the sampling are taken into account. A large number of model realizations are created using photometry, background CMDs, and artificial stars.  Then the SFH is recalculated for each realization. The variance of these SFHs then provides the uncertainty on the best-fitting SFH assigned to each location \citep{Dolphin2012,Dolphin2013}.

The SFHs of these regions are consistent with the existence of an RSG candidate in our selection region. We tested this consistency by generating many model CMDs in the IR bands using a sample of our SFHs and calculating the fraction of draws from the model that contained at least one RSG candidate.  All SFH models had at least 25\% probability of producing a star with IR color and magnitude inside of our selection region, and 50\% had at least 95\% probability of producing such a star.

\subsection{RSG evolutionary models} \label{models}
In this section, we briefly describe the models used to compare to our age measurements and use the evolutionary tracks and isochrones to test their consistency with our age measurements. 

\subsubsection{PARSEC and Geneva} \label{p&g_model_info}
Two widely used sets of single stellar evolution models are from the PARSEC and Geneva \citep{Ekstrom2012} groups.  We compare our age results against each of these sets below.

Figure \ref{fig:ISO_cmd} contains a plot that shows the evolutionary tracks of stars with a few different masses entering our defined RSG space on the left panels. The colored regions indicate the area covered by our age ranges of interest on both the track plot and the CMD plot. These areas were determined from the magnitude values given in the isochrones by first removing points that do not pass our luminosity, effective temperature, magnitude, and color limits. We then generated a polygon to cover the remaining points. Thus, the polygons in the CMD correspond to the polygons of the same color in luminosity and temperature space in the model isochrones and evolutionary tracks.

We applied the same process to the Geneva \citep{Ekstrom2012} tracks and isochrones. The Geneva model has fewer data points in their model grids. We use isochrone files with a log age of 6.75-7.6 in increments of 0.05 log age. We use tracks from stars with a mass of 10, 12, 15, 20, 25, and 32 solar masses.

\subsubsection{BPASS} \label{b_model_info}
\begin{figure}[!htb]
    \includegraphics[width=\textwidth]{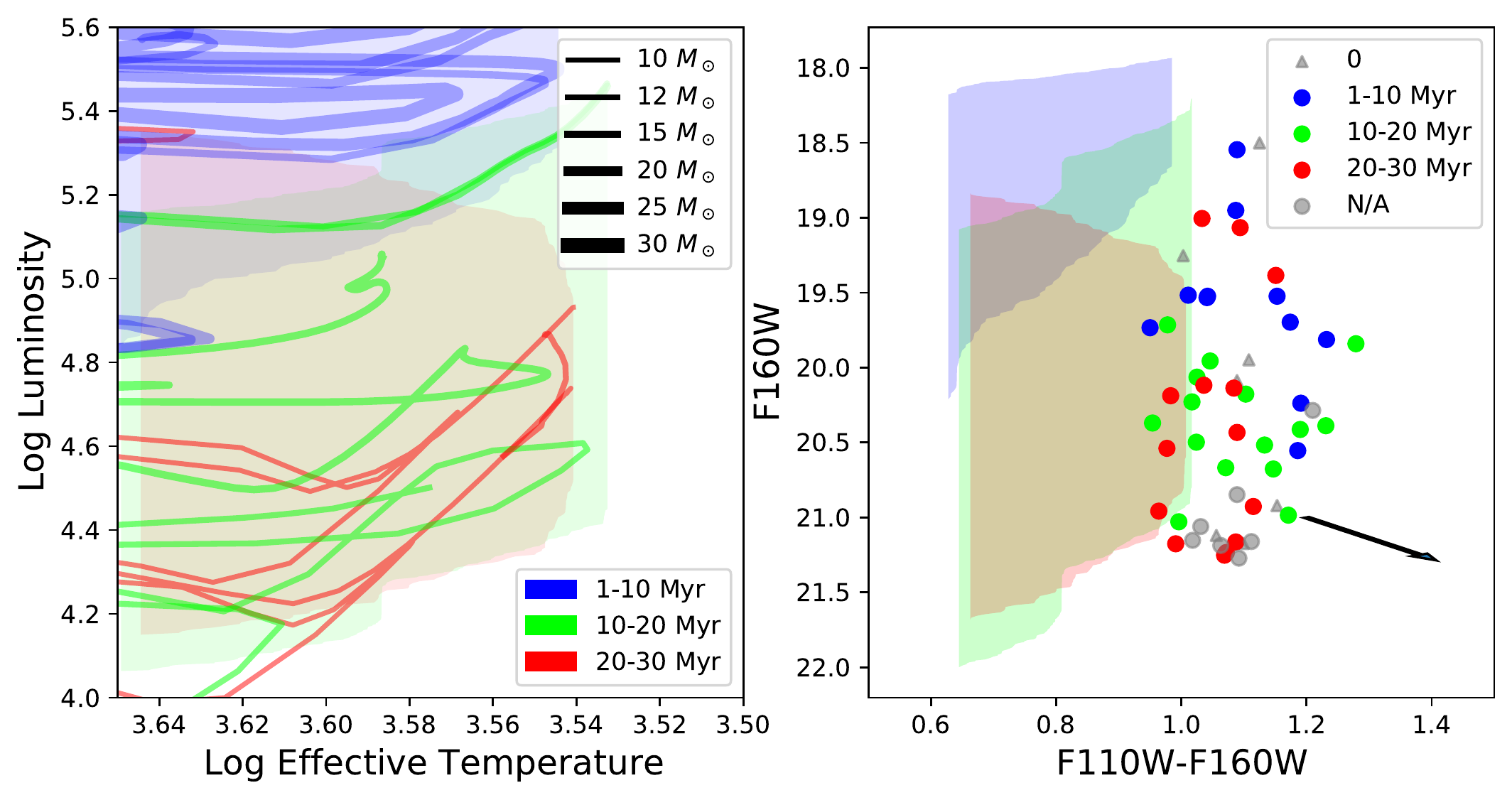}
    \includegraphics[width=\textwidth]{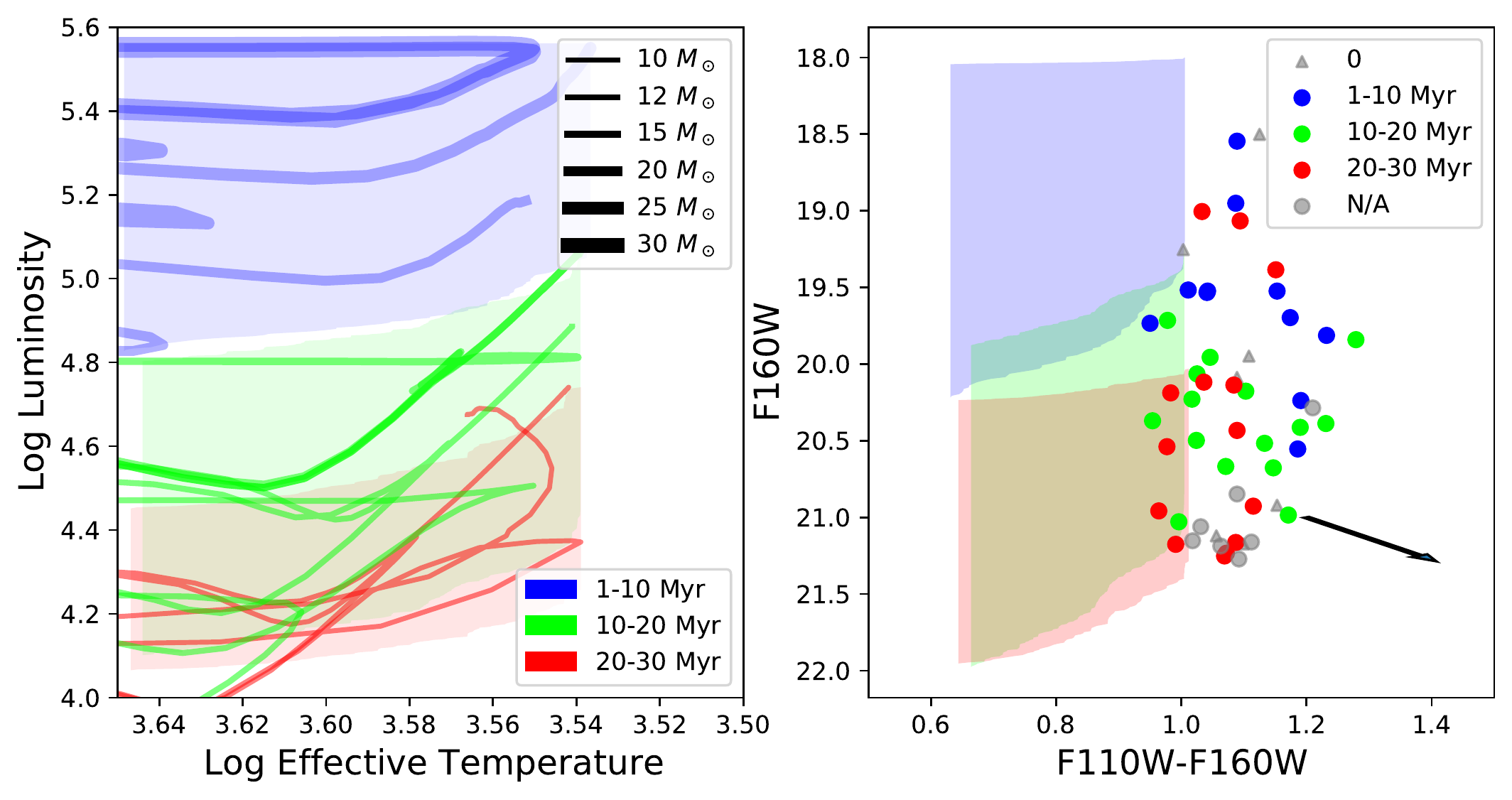}
    \caption{\label{fig:ISO_cmd_BPass} These plots are the same as the previous evolutionary track and CMD plots but with the BPASS Binary (top) and Secular (bottom) model sets. As the legend suggests, the thickness of each line indicates the associated mass. The 1-10 Myr tracks have a lower alpha in order to see some of the overlap between different mass values. The BPASS models have too many tracks for each mass to plot in an orderly manner. Instead, we selected a high luminosity track, a low luminosity track, and an average luminosity track to the plot which effectively represented the range of values in the models. We see in the BPASS models general agreement with our results, similar to that of the PARSEC models. However, the binary model set shows the F160W magnitude range of older stars extending up to the 19th magnitude. This indicates we could possibly have binary RSGs in our sample.}
 \end{figure}

The BPASS models \citep{Eldridge2017,StanwayandEldridge2018} are the final set of models we compared against our results. The BPASS models have both binary and secular options. As a result, the process of creating the figures containing the data from BPASS was quite different from both the previous models. Instead of separate files for tracks and isochrones, BPASS comes in one large suite of stellar models, including a couple thousand individual binary models. These models were labeled by their metallicity, mass of the initially less massive star (treated as primary in subsequent evolution), mass of the remnant after death of the initially more massive star, and the log of the binary period after remnant formation. For simplicity, we only considered the models with a solar metallicity and 10, 12, 15, 20, 25 and 30 solar masses, leaving a grid of close to 200 models for each mass for comparison against our RSG observations. Instead of plotting all 200 for each mass, in Figure \ref{fig:ISO_cmd_BPass} we plot a file with a relatively high luminosity maximum for the mass, a file with a relatively low luminosity minimum, and then an average luminosity file. We found this technique captured the range of temperature and luminosities we would find with plotting all the files but is much easier to see. The resulting regions of temperature, as well as CMD space, for each age are plotted in Figure \ref{fig:ISO_cmd_BPass} using the same color coding as for the single-star PARSEC and Geneva models. 

\section{Results and Discussion} \label{results}
The results from MATCH are shown in Table \ref{tab:MATCH_results}. Column 1 has the unique identifier for each RSG candidate. Columns 2 and 3 contain the best fit $dA_{v}$ and $A_{v}$ values. Column 4 has the best age result in Myr. Finally, column 5 contains a flag used to indicate the quality of the candidate as well as which sample the candidate belongs to. A means the candidate has shown only signs of being a quality candidate. B indicates the candidate is in a potential binary system, C means the star is a potential low-mass contaminant (AGB star), D means the candidate location was analyzed with the F435W and F606W filter combination, F means the star is a potential foreground star, M indicates the candidates that were our most likely mismatches, and IN means the location had insufficient coverage in the surrounding area to get an age estimate from MATCH. A flag of 1 means the candidate belongs to the primary sample comprised of only single matches between HST and Spitzer sources while 2 means the candidate belongs to the secondary sample of candidates that had multiple matches between HST and Spitzer sources.

\begin{figure}[!htb]
    \center{\includegraphics[width=.5\textwidth]{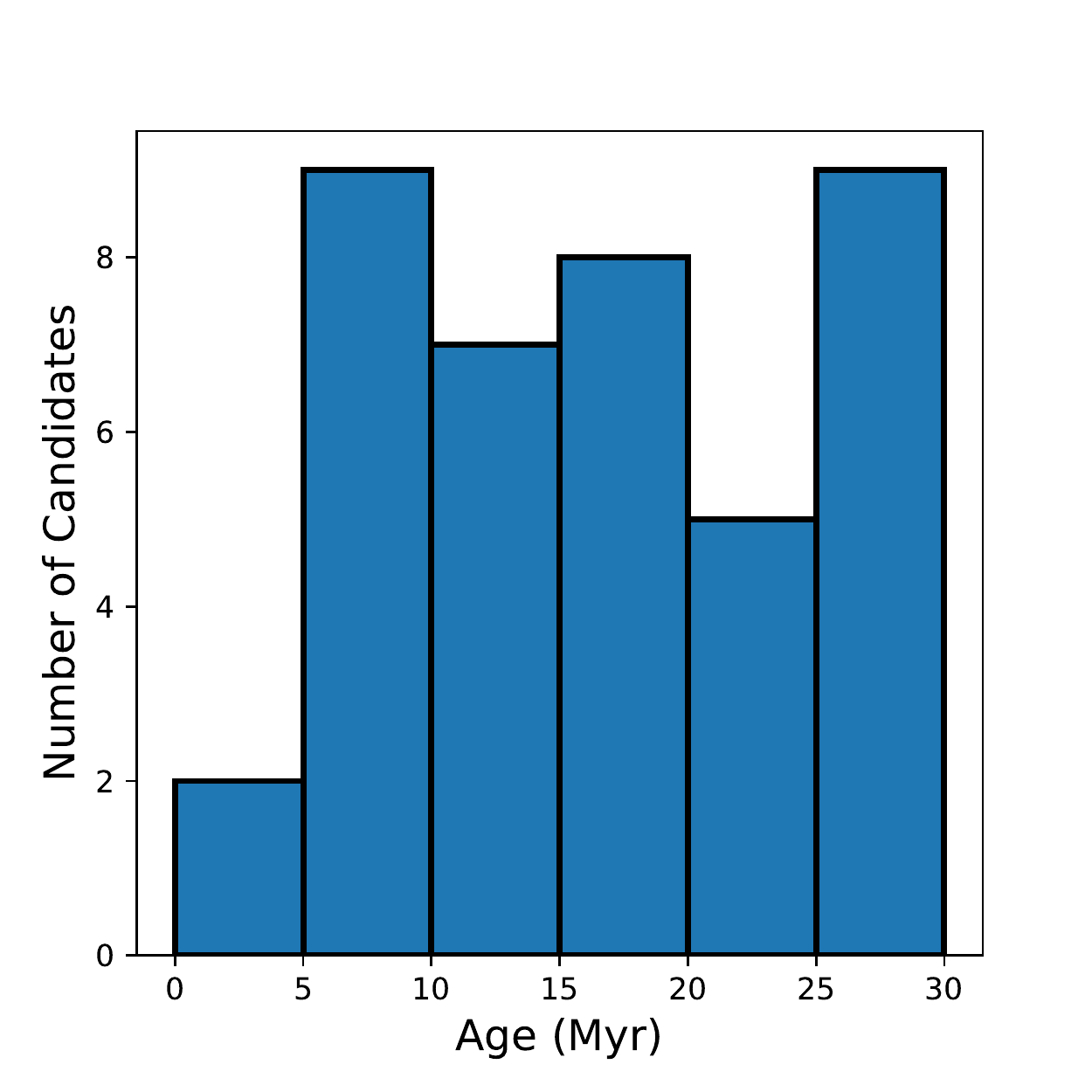}}
    \caption{\label{fig:Compare} A histogram showing the age results from MATCH. Almost all candidates are older than 5 Myr, as may be expected from a standard initial mass function.}
 \end{figure}

Figure \ref{fig:Compare} shows a histogram of the MATCH results with contamination CMDs. The figure shows nearly all candidates are older than 5 Myr, as one might expect given the small number of very massive stars produced by a standard initial mass function. We found seven RSG candidate locations showing no significant star formation $\leq{30}$ Myr, which indicates that they are either not RSGs or have unusually high velocities that have moved away from their siblings. For example, these stars could be AGB or post-AGB contaminants that entered our sample. AGB stars are older than our age cap of 30 Myr and would receive an age value of zero from MATCH, as shown in Figure \ref{fig:ISO_cmd}. These potential contaminants were excluded from the histogram in Figure \ref{fig:Compare}

\begin{figure}[!htb]
    \includegraphics[width=.5\textwidth]{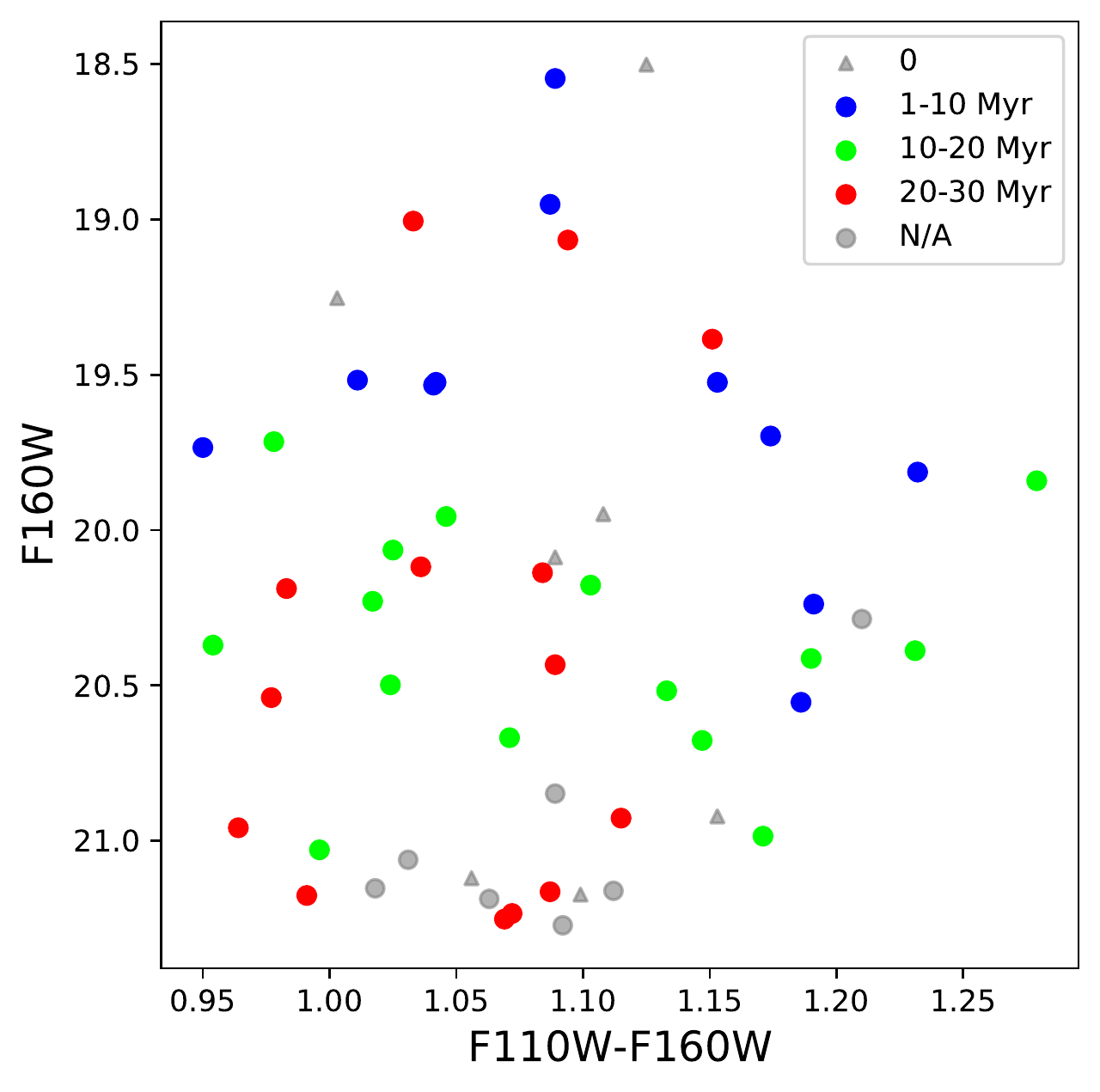}
    \includegraphics[width=.5\textwidth]{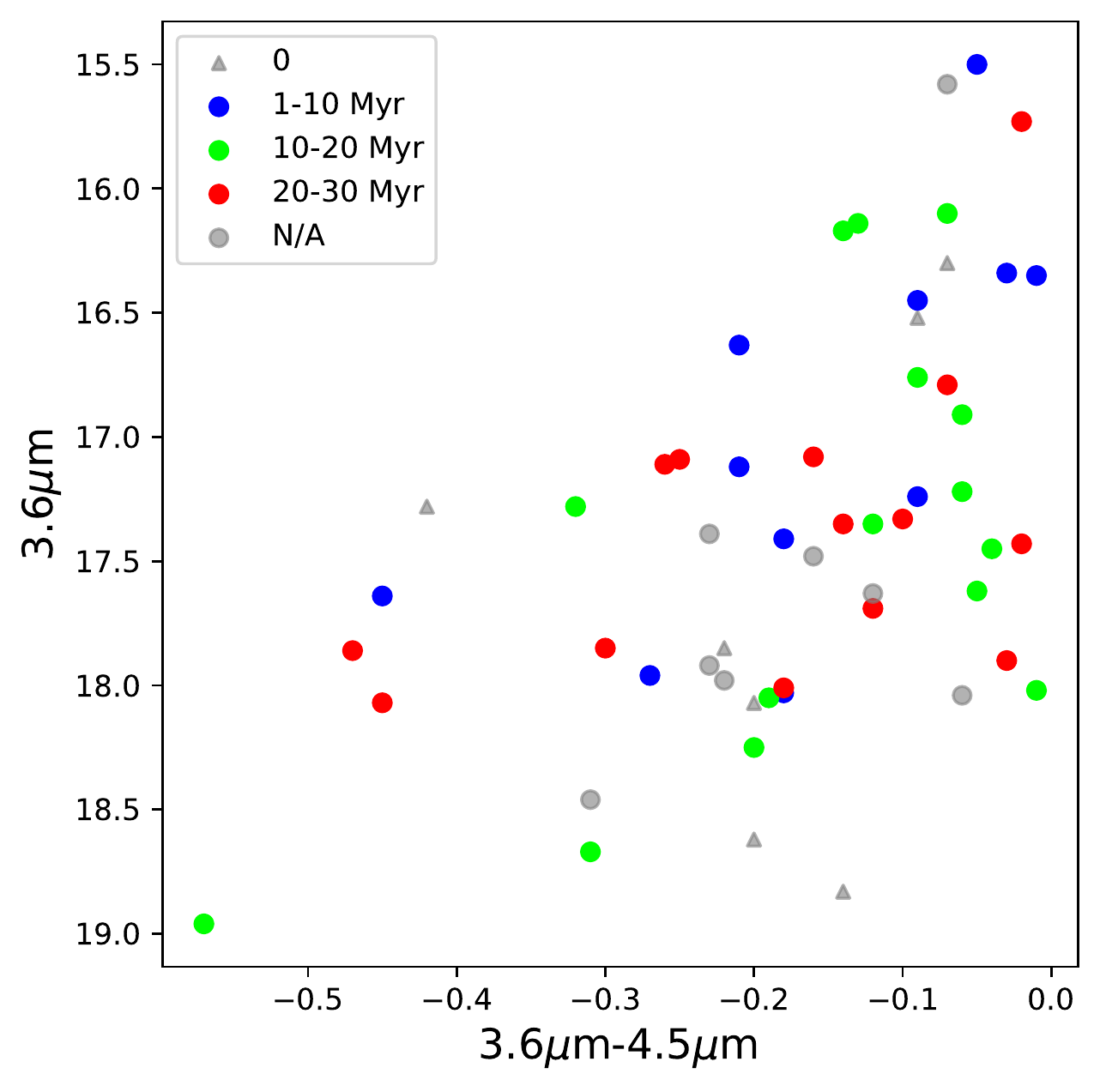}
    \caption{\label{fig:CMD_age} The left panel is a CMD in \emph{HST} filters of the 40 primary sample candidates color coded by their age. The y-axis is the magnitude of the candidates in F160W and the x-axis is the F110W-F160W color of each candidate. The right panel is the same CMD using Spitzer filters with the 3.6$\mu$m mag on the y-axis and the 3.6$\mu$m-4.5$\mu$m color on the x-axis.}
 \end{figure}

\begin{figure}[!htb]
    \center{\includegraphics[width=\textwidth]{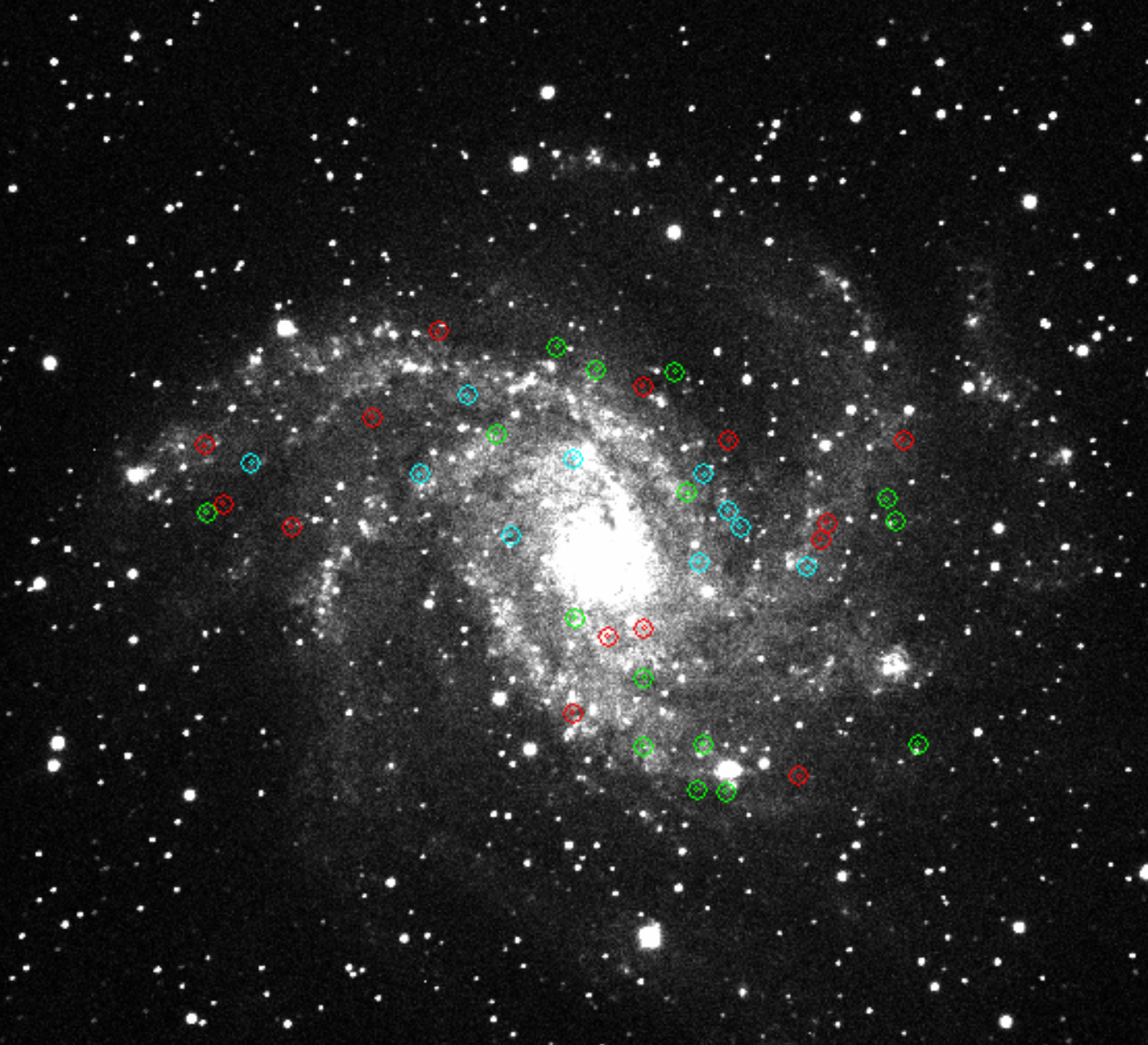}}
    \caption{\label{fig:galaxy} The same as Figure \ref{fig:IRgalaxy} but overlaid with the locations of the primary sample of RSG candidates, which are color coded by age.
     Blue: 1-10 Myr,
     Green: 10-20 Myr,
     Red: 20-30 Myr.
     North is up and east is to the left. The radius of each circle is 5".}
 \end{figure}

\begin{figure}
    \centering
    \begin{minipage}{.3\textwidth}
        \centering
        \includegraphics[width=\textwidth]{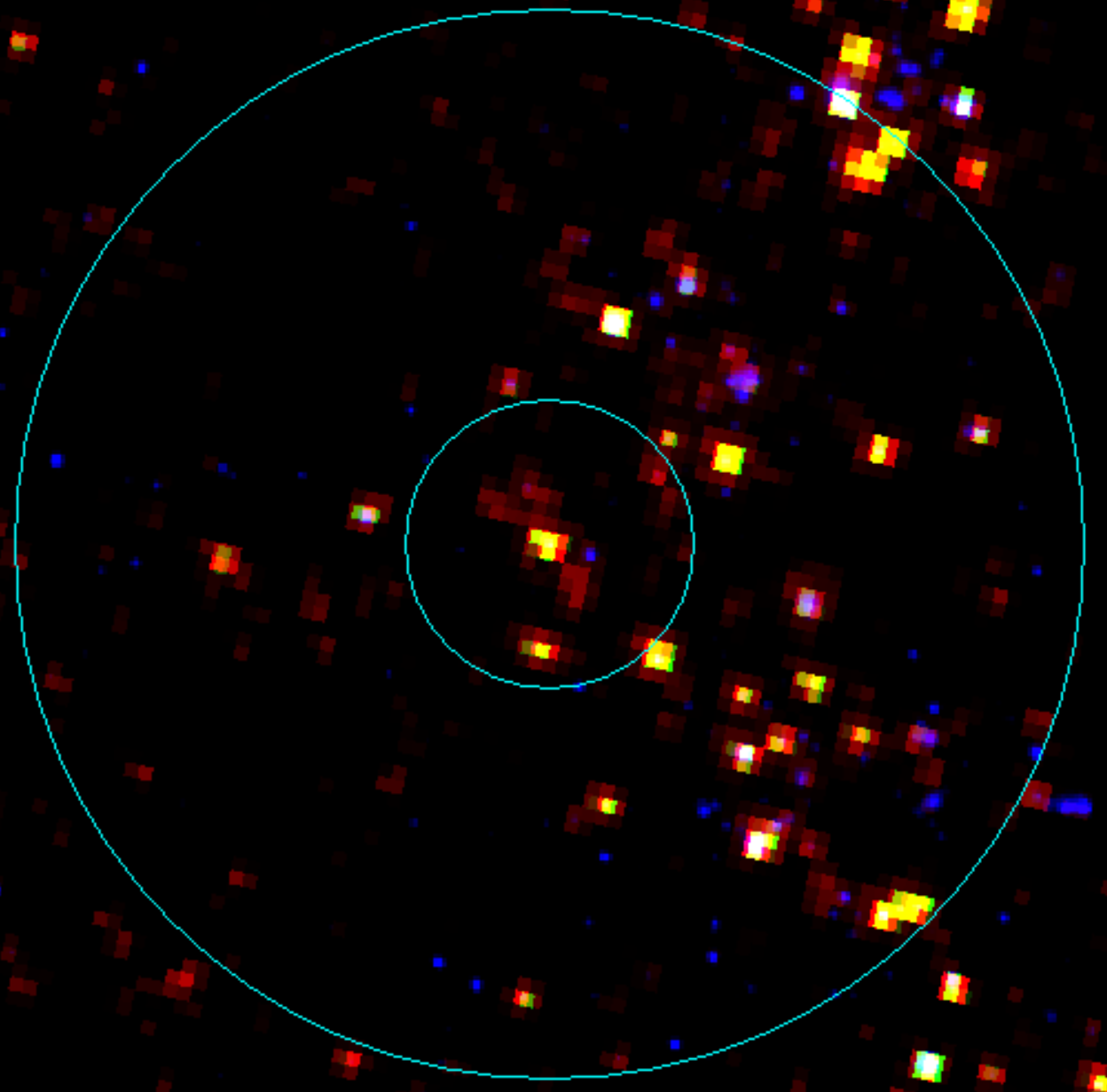}
        \\ RSG 767 - 9 Myr
    \end{minipage}
    \begin{minipage}{.3\textwidth}
        \centering
        \includegraphics[width=\textwidth]{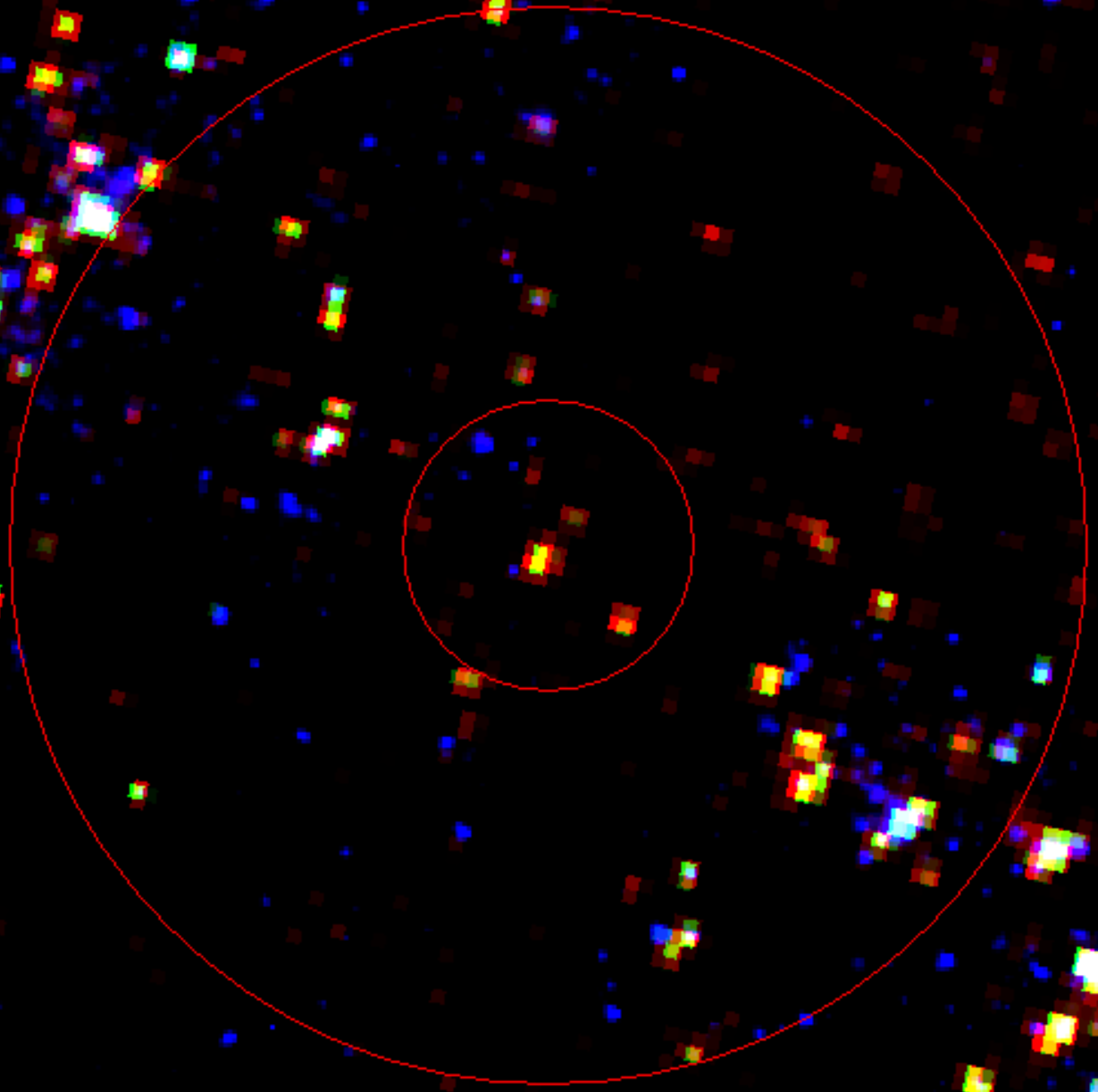}
        \\ RSG 928 - 24 Myr
    \end{minipage}
    \begin{minipage}{.3\textwidth}
        \centering
        \includegraphics[width=\textwidth]{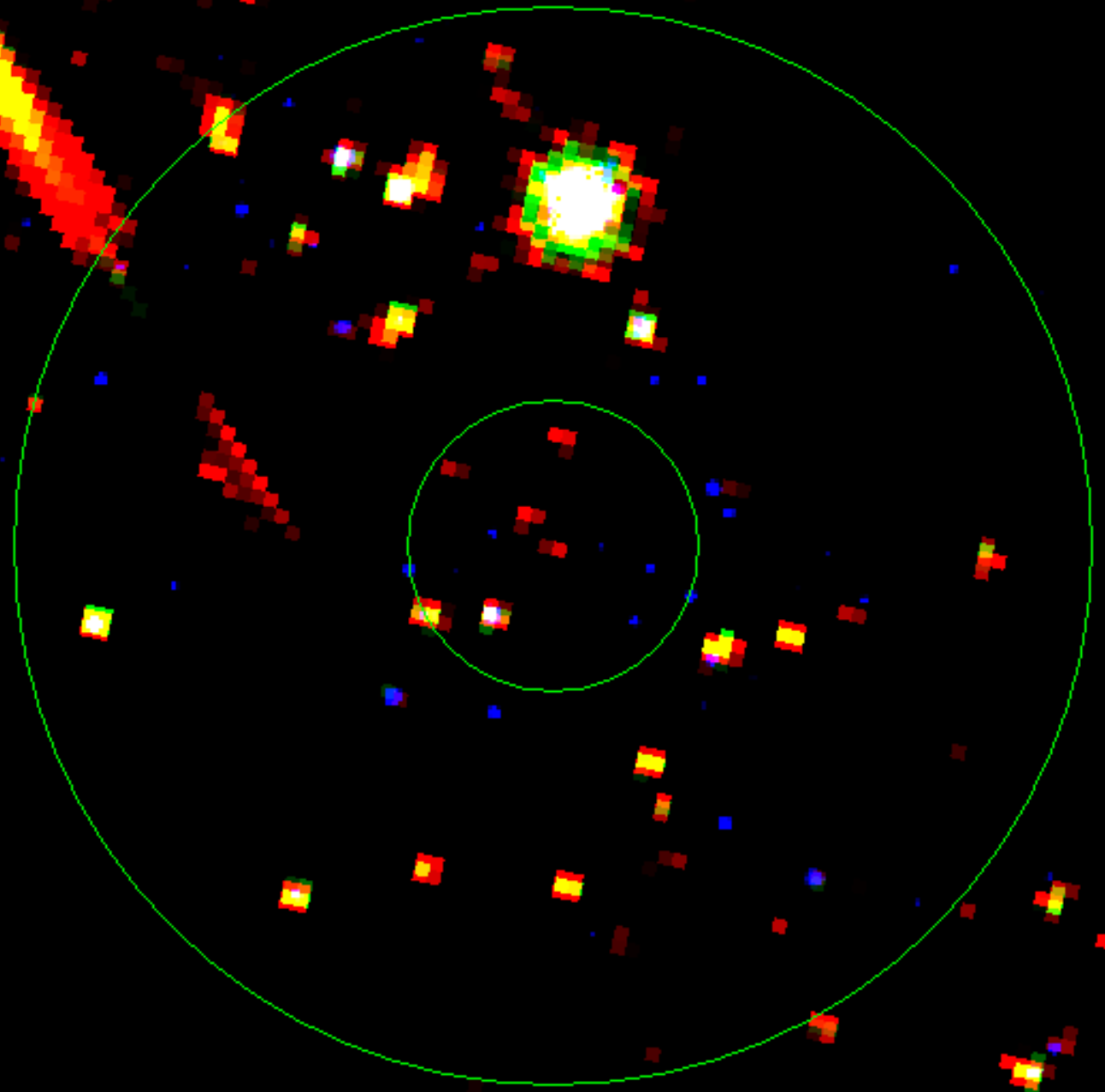}
        \\ RSG 1282 - 14 Myr
    \end{minipage}
    \begin{minipage}{.3\textwidth}
        \centering
        \includegraphics[width=\textwidth]{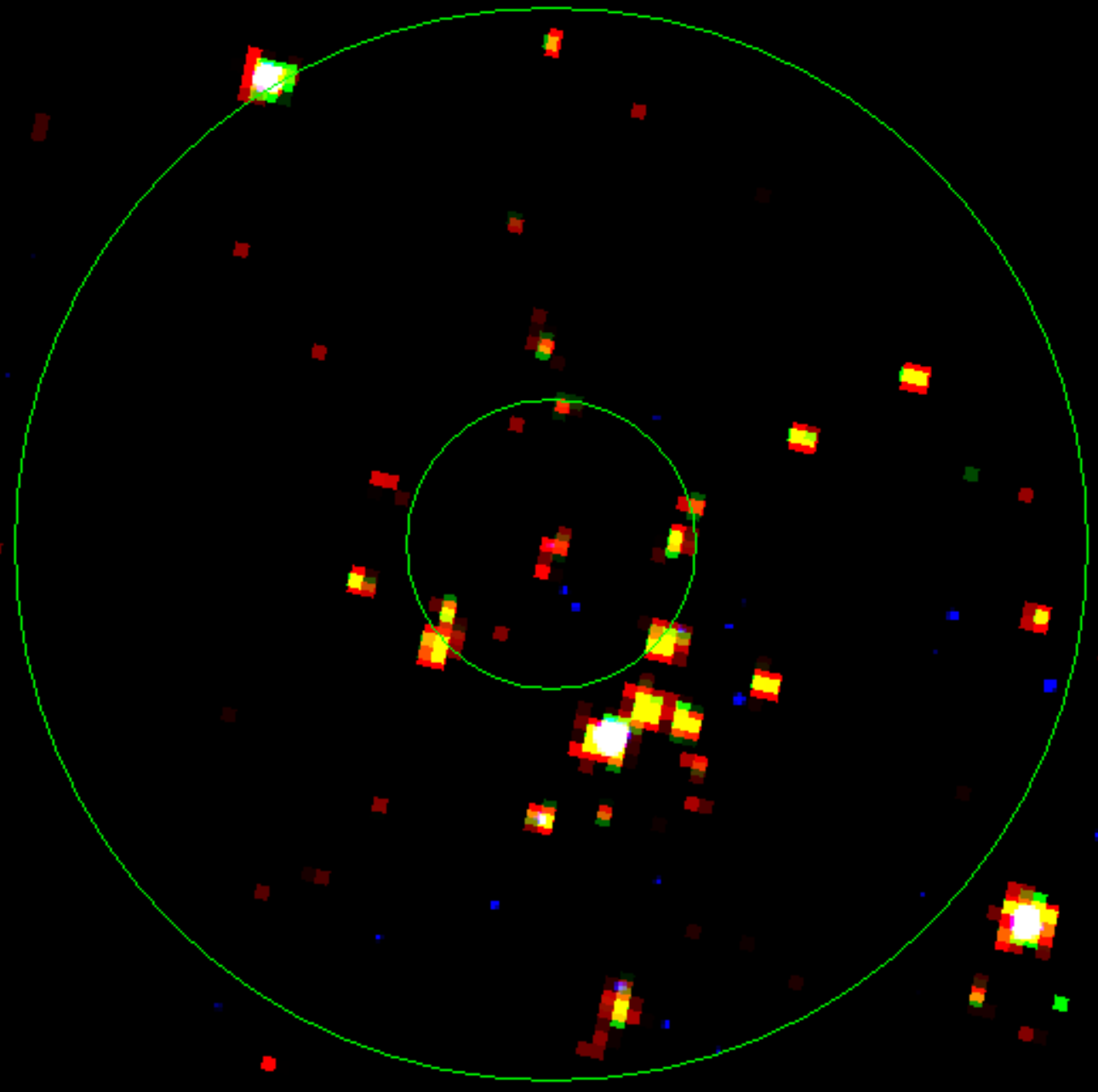}
        \\ RSG 1321 - 15 Myr
    \end{minipage}
    \begin{minipage}{.3\textwidth}
        \centering
        \includegraphics[width=\textwidth]{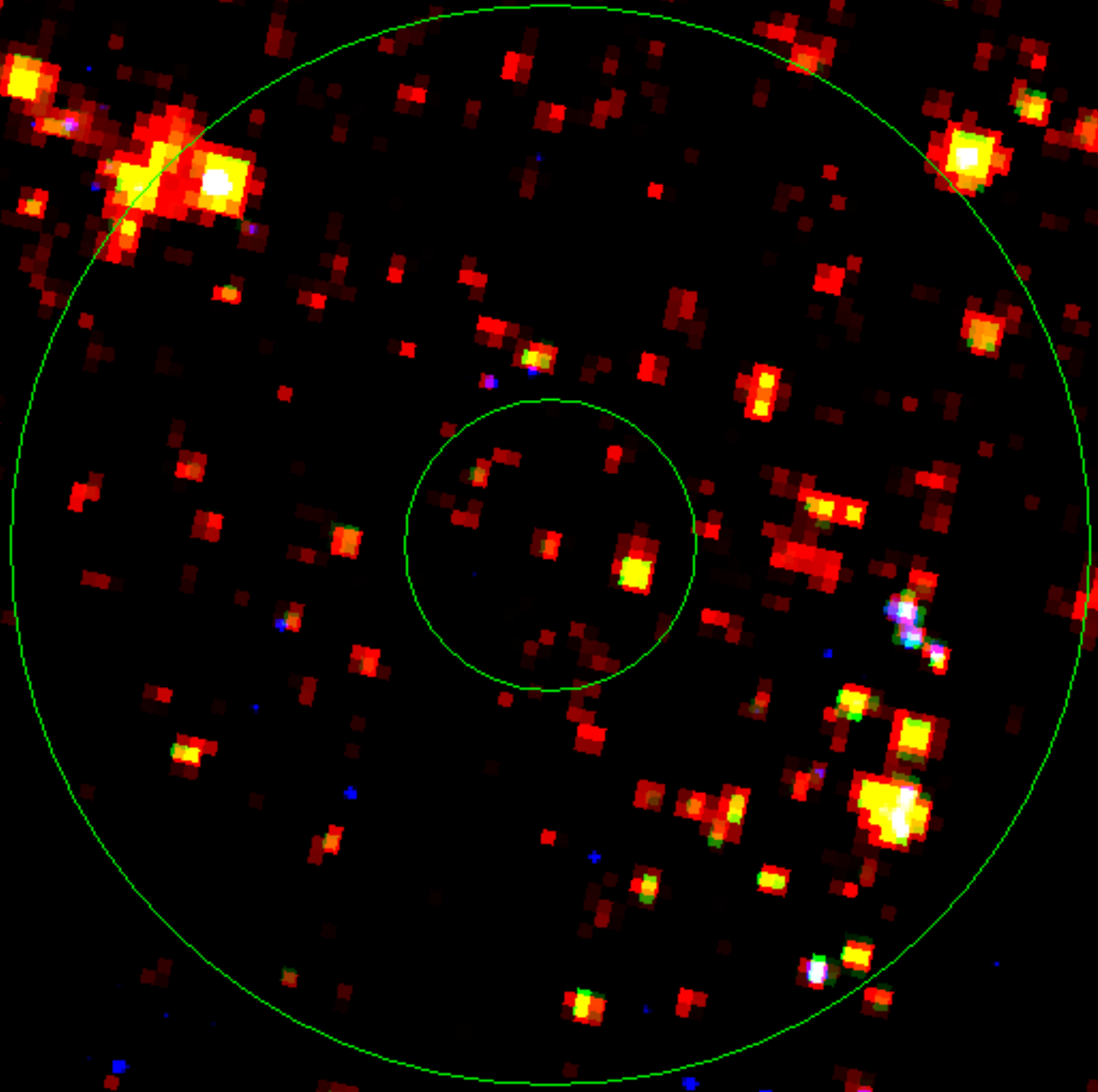}
        \\ RSG 1355 - 10 Myr
    \end{minipage}
    \begin{minipage}{.3\textwidth}
        \centering
        \includegraphics[width=\textwidth]{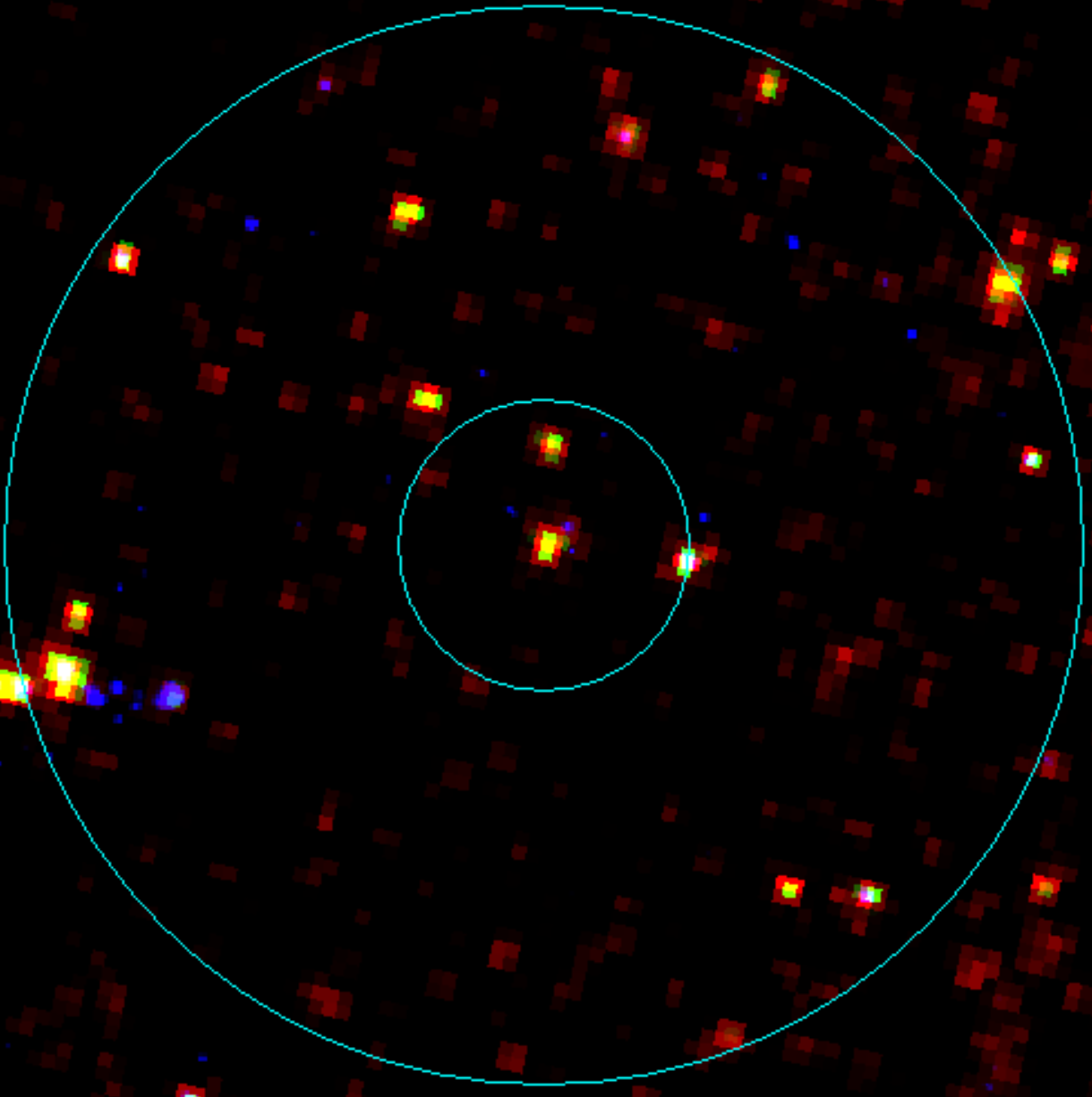}
        \\ RSG 1534 - 4 Myr
    \end{minipage}
    \begin{minipage}{.3\textwidth}
        \centering
        \includegraphics[width=\textwidth]{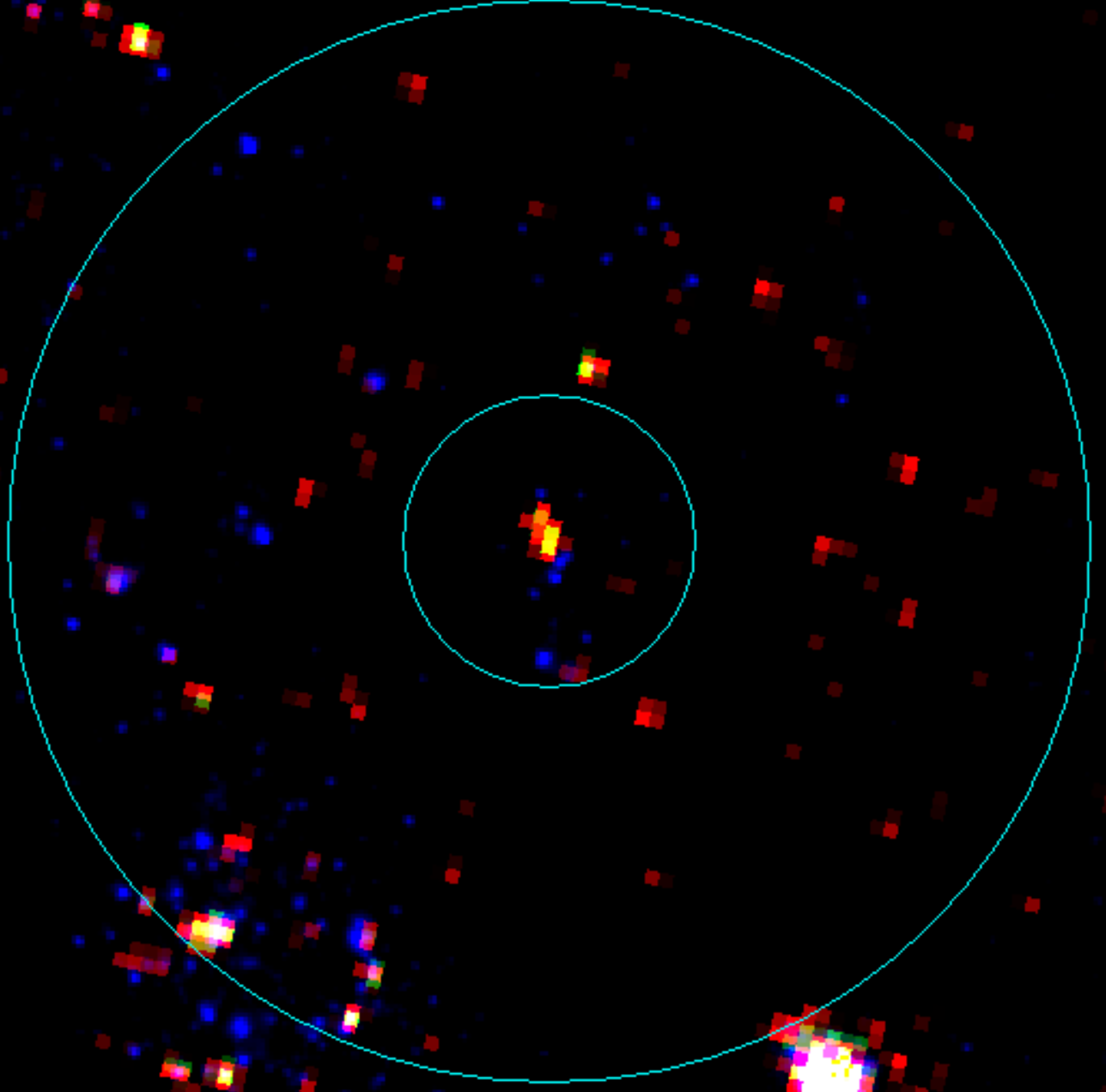}
        \\ RSG 1539 - 5 Myr
    \end{minipage}
    \begin{minipage}{.3\textwidth}
        \centering
        \includegraphics[width=\textwidth]{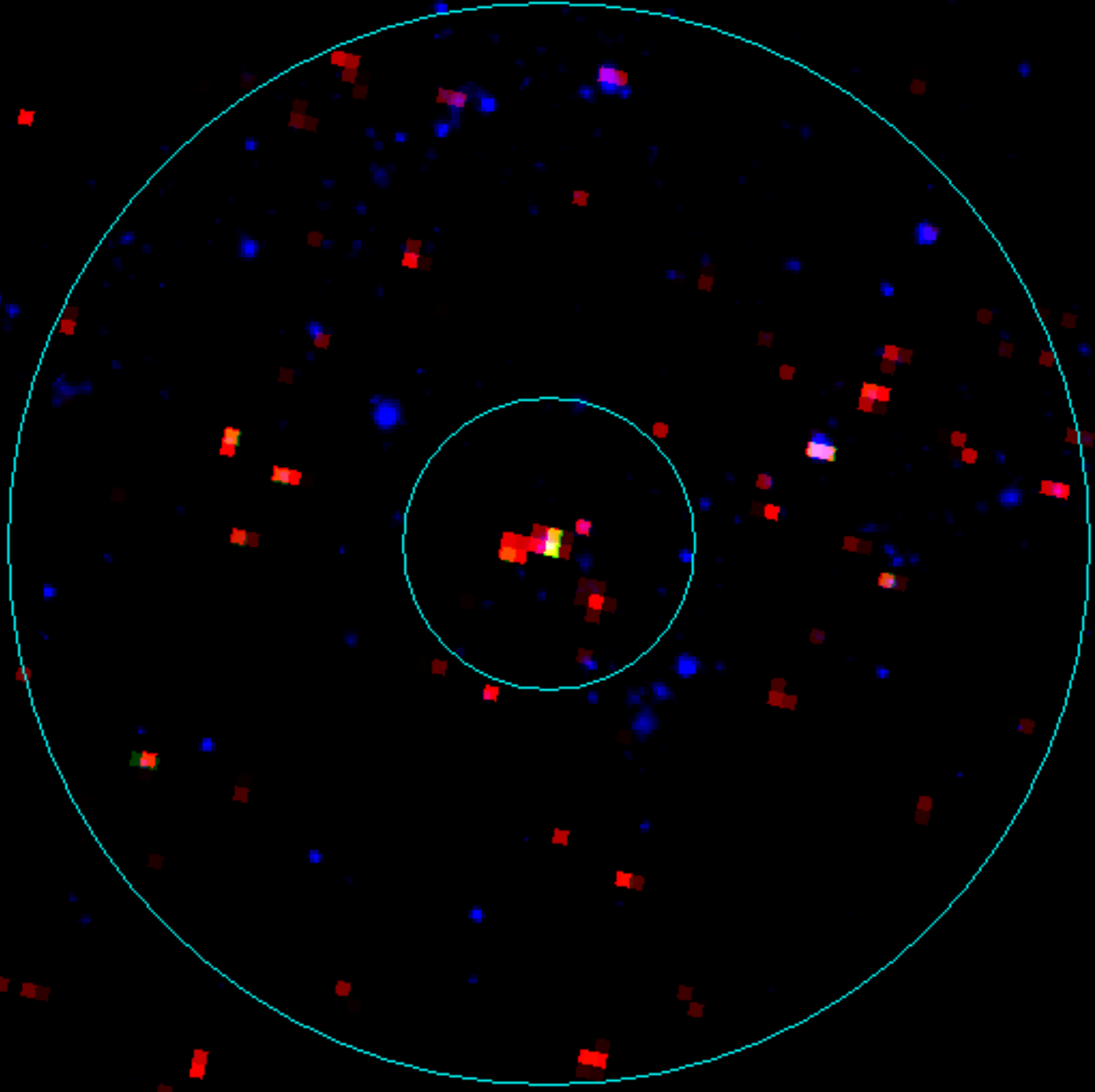}
        \\ RSG 1661 - 5 Myr
    \end{minipage}
    \begin{minipage}{.3\textwidth}
        \centering
        \includegraphics[width=\textwidth]{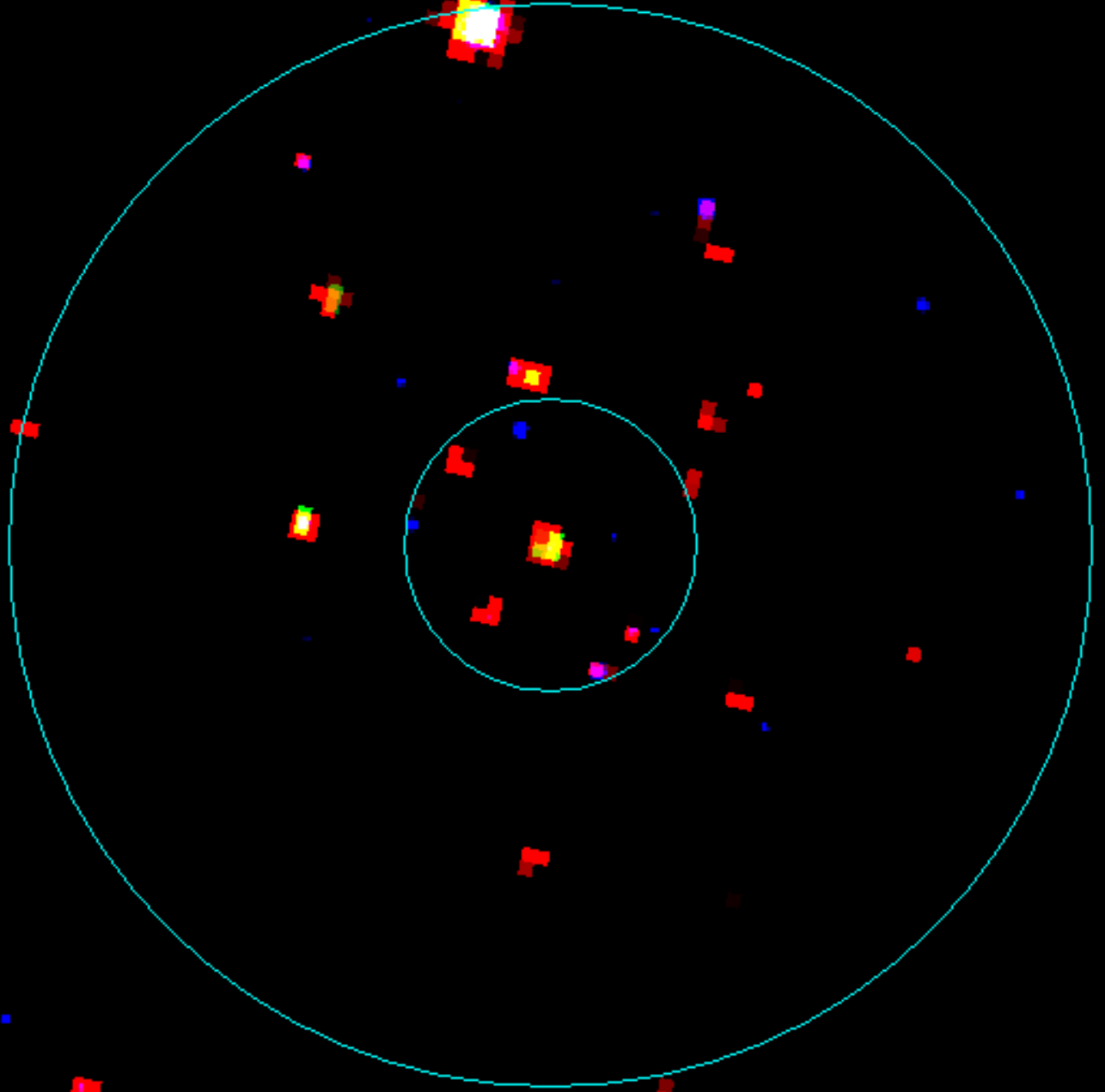}
        \\ RSG 1966 - 5 Myr
    \end{minipage}
    \caption{These are RGB snapshots of each RSG candidate location color coded by their age grouping. In the image, F160W is the red filter, F110W the green, and F606W the blue. The images are 10" by 10" with a small inner circle of radius 1".39 (50 pc). Each snapshot has the R.A., decl., and age of the candidate. \label{fig:stamps}}
\end{figure}

A pair of CMDs in the \emph{HST} filters and Spitzer filters is shown in Figure \ref{fig:CMD_age} showing the 40 candidates plus the seven candidates returning an age of zero and seven candidates that did not have sufficient detected stars in the surrounding area to calculate an age. Figure \ref{fig:galaxy} shows NGC~6946 overlaid with the locations of the RSG candidates color coded by age. Finally, we show close-up color images for some RSG candidates, in Figure \ref{fig:stamps}. All of the primary sample candidates with an 'A' flag will have images in the online journal as stated in Figure \ref{fig:fset}. These images use the F160W filter for red, F110W for green, and F606W for blue. 
 
\subsection{Comparing Our Measured Ages to Models} \label{compare}
In this section we discuss the different models used to interpret our results and the techniques for doing so. We also discuss how AGB stars could show up in the models and what that means for our candidate selection. Without spectroscopy, we cannot measure a mass for these candidates, but in this section we use the models to provide mass estimates for the candidates that are most likely to be RSG stars. 


\subsubsection{PARSEC and Geneva} \label{p&g_models}
\begin{figure}[!htb]
    \center{\includegraphics[width=\textwidth]{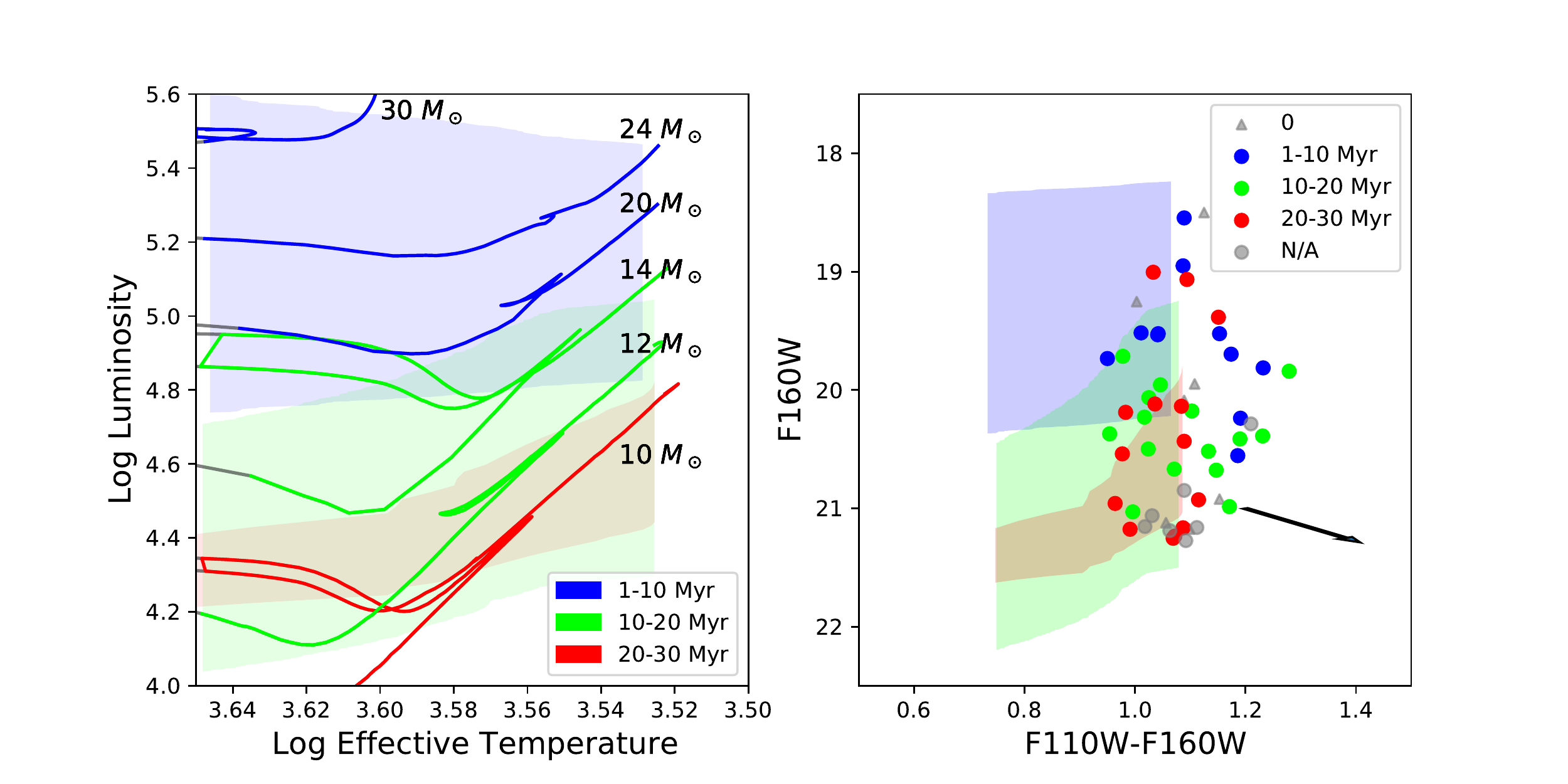}}
    \begin{minipage}{.49\textwidth}
        \centering
        \includegraphics[width=\textwidth]{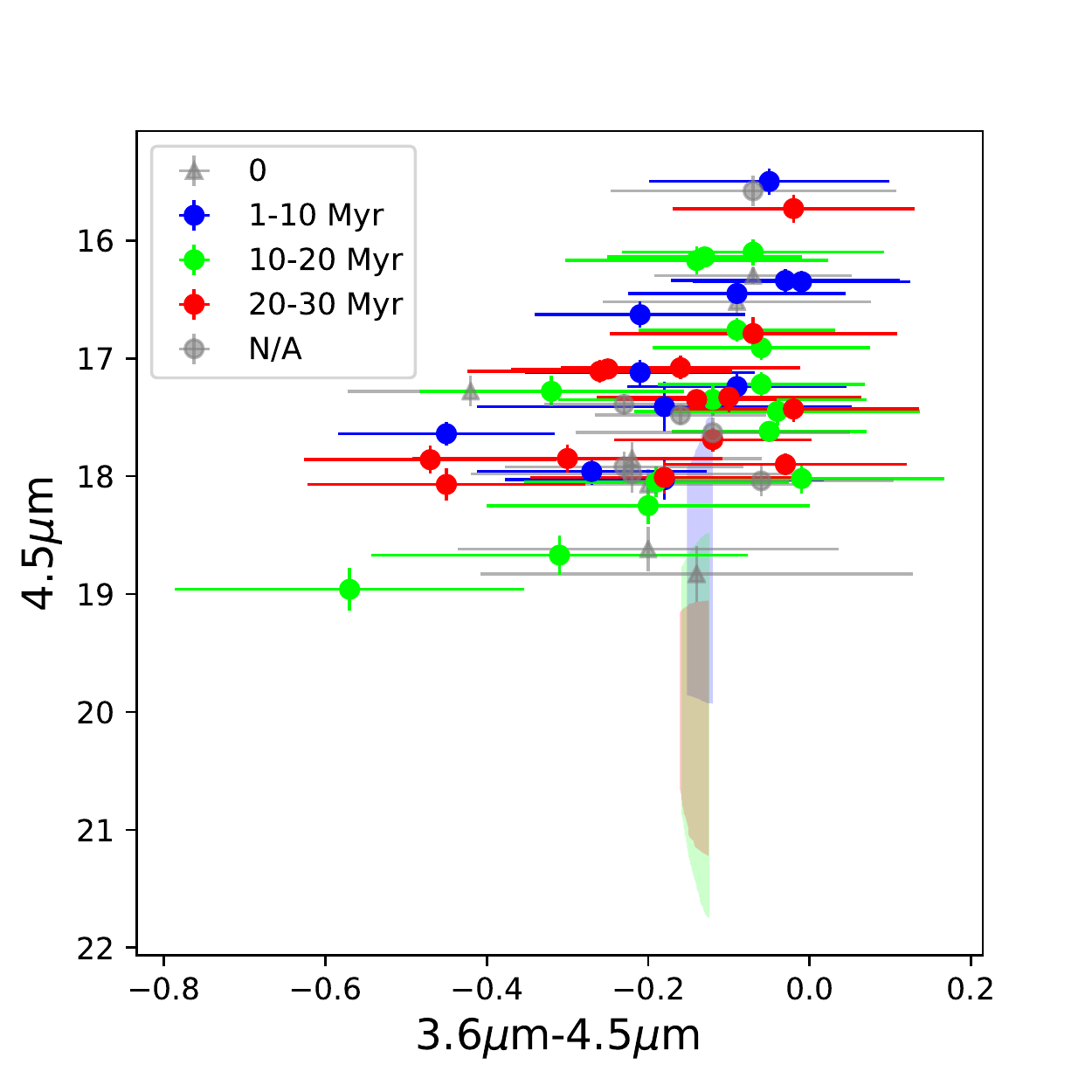}
    \end{minipage}
    \begin{minipage}{.49\textwidth}
        \centering
        \includegraphics[width=\textwidth]{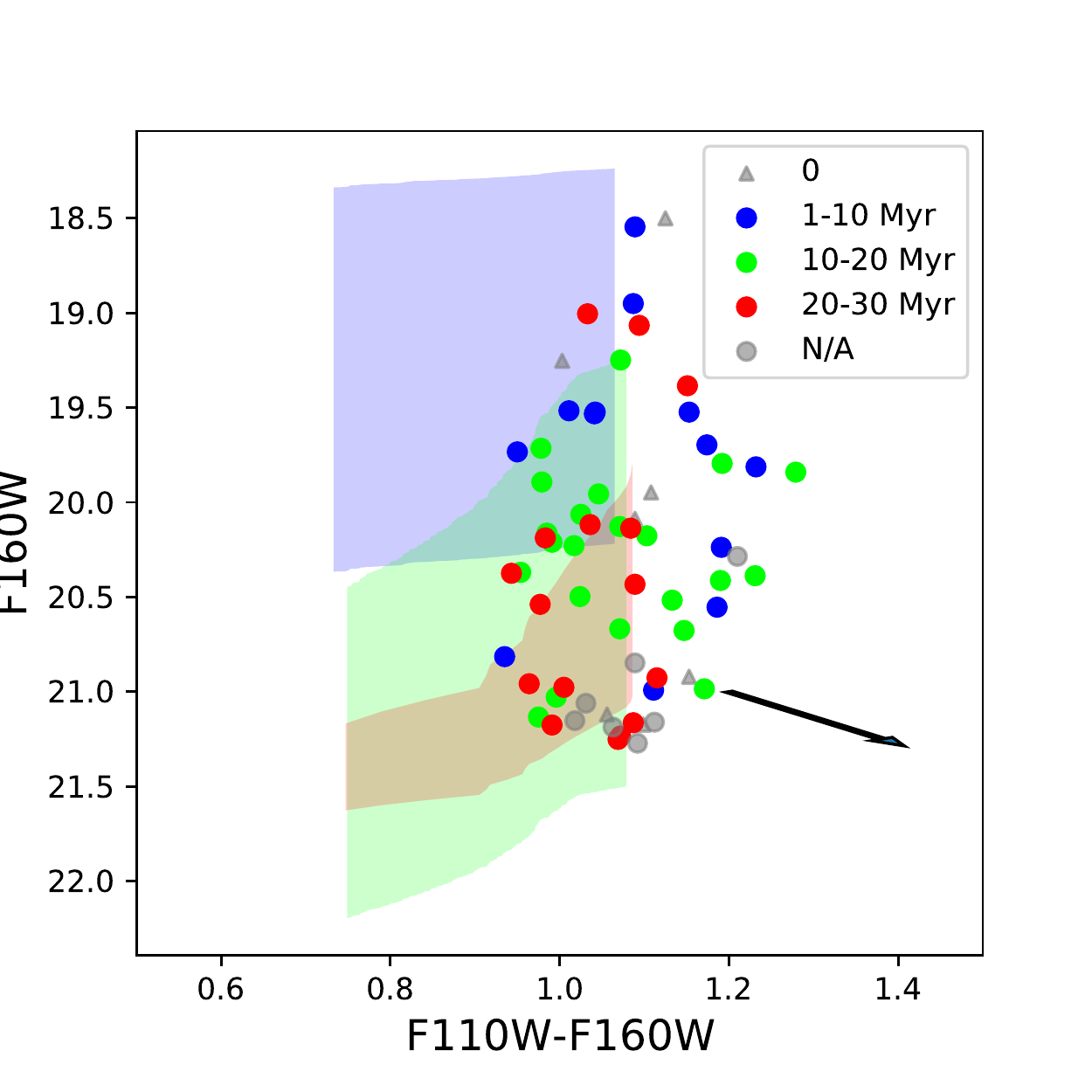}
    \end{minipage}
    \caption{\label{fig:ISO_cmd} The plot on the top left shows evolutionary tracks from the PARSEC models of different ages with axis limits isolating the RSG space defined by effective temperature and luminosity limits. Blue represents ages between 1 and 10 Myr, green is between 10 and 20 Myr, and red is between 20 and 30 Myr. There are also regions of color that represent the ranges of allowed effective temperature and luminosity values allowed in the models. On the top right is the CMD of our candidates with similar regions created using allowed magnitude and color values from the models. A description of how these regions are created is given in Section \ref{p&g_model_info}. The arrow represents a reddening vector with F160 reddened by 0.25 mag. The regions and CMD points follow the same color coding scheme as before. The CMD includes smaller gray triangles that represent the seven candidates that showed no significant star formation younger than 30 Myr and gray points that represent the candidates with insufficient coverage to be analyzed by MATCH. The bottom left panel shows a similar plot but instead uses Spitzer filters in both the plotting of the points (with error bars) and drawing the regions. The modeled Spitzer magnitudes and colors show a very narrow range of accepted color values which is not represented in the data at all. This discrepancy might be able to be described by dust emission. The bottom right panel shows the same CMD as the top right panel but includes all 51 potential candidates, instead of just the primary sample, plus the seven stars that returned an age of zero from MATCH and the seven candidates that did not have sufficient detected stars in the surrounding area to calculate an age (denoted by N/A). The modeled \emph{HST} regions match up much more closely with our results as compared to the modeled Spitzer regions.}
 \end{figure}

Figure \ref{fig:ISO_cmd} shows the red supergiant region of the evolutionary tracks from the PARSEC models along with the age range that corresponds with the stars that passed the cuts. On this plot we also show polygons of parameter space on the CMD color coded by the age bin. These polygons are the ranges of F160W and F110W-F160W allowed values from the PARSEC isochrones. In Figure \ref{fig:ISO_cmd}, the panels on the right-hand side represent the models and the RSG candidates in the \emph{HST} filters while the bottom panel on the left shows the models and candidates in Spitzer filters. 

We also used the PARSEC models to investigate how AGB stars could potentially contaminate our sample. Figure \ref{fig:possAGB} is a plot showing how AGB stars could intrude on our defined space. Lower-mass stars from 4-9\(M_\odot\) are plotted along with the original selection from the \emph{HST} part of Figure \ref{fig:ISO_cmd}. The luminosity minimum was lowered from 4.5 to 4.2. This extended the allowed region fainter than our RSG selection would allow. One important note to consider is that the 4-9\(M_\odot\) evolutionary tracks are all older than our 30 Myr maximum allowed age in the MATCH analysis.

\begin{figure}[!htb]
    \center{\includegraphics[width=\textwidth]{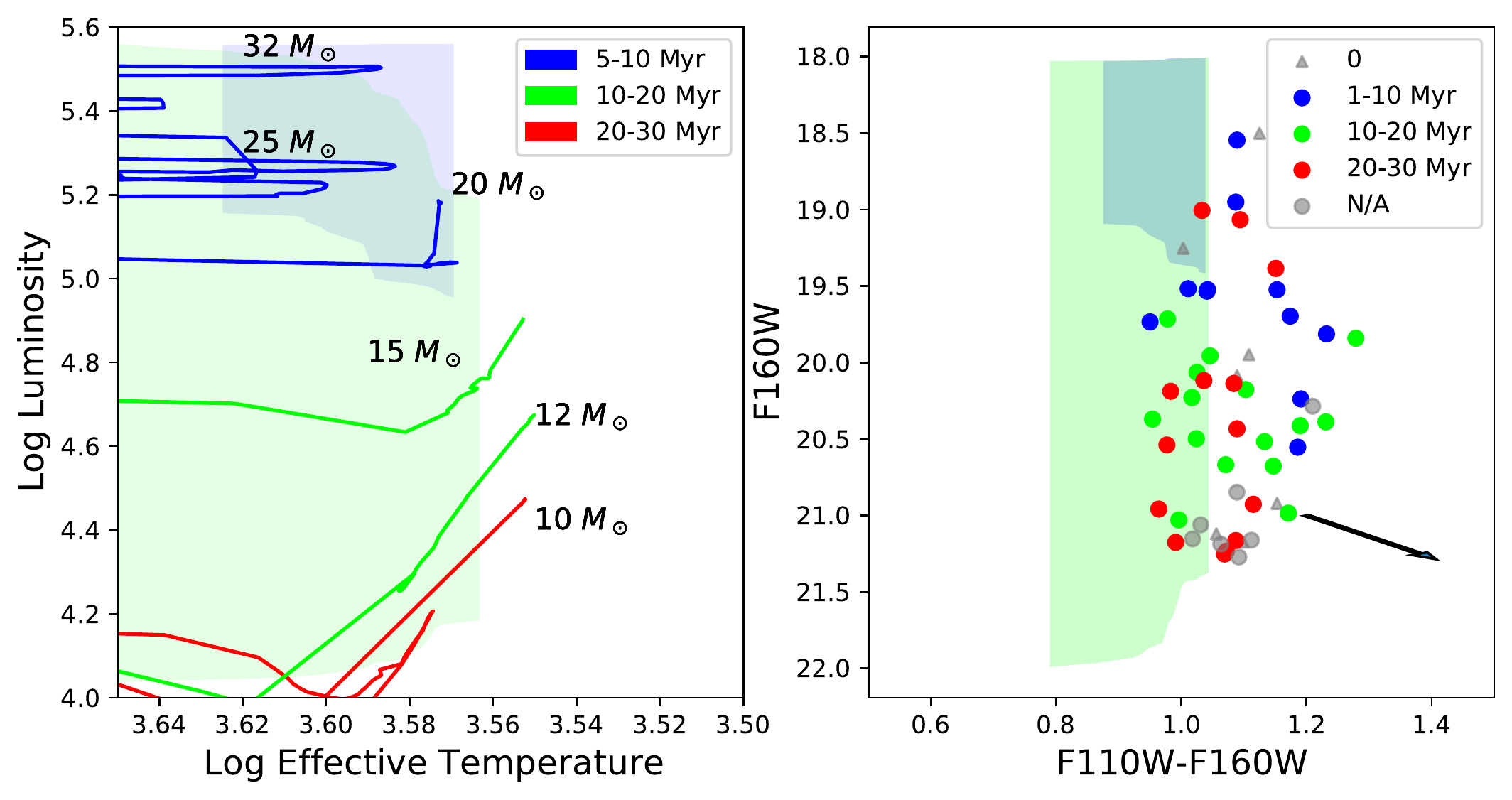}}
    \caption{\label{fig:ISO_cmd_geneva} These plots are the same as the PARSEC plots above but are with the Geneva model instead. The plot on the left shows evolutionary tracks from the Geneva models of different ages with axis limits isolating the RSG space. There are also regions of color that represent the ranges of allowed effective temperature and luminosity values allowed in the models. The Geneva models did not have stars younger than 5 Myr enter this defined effective temperature and luminosity space. On the right is the CMD of our candidates with similar regions created using allowed magnitude and color values from the models.}
 \end{figure}

Figure \ref{fig:ISO_cmd_geneva} shows the red supergiant portions of the Geneva models for direct comparison to PARSEC. There were no models in the 20-30 Myr age range with temperatures and luminosities that would place them in our photometry sample, so no red region is drawn on the CMD. The 10-20 Myr range dominated the entire color-magnitude region covered by our RSG candidates. By using multiple single-star evolution models we can show the discrepancies in the models in this phase of evolution. Between PARSEC and Geneva we show differences in the range of luminosity and temperature according to the age of RSGs.

\subsubsection{BPASS} \label{b_models}
Figure \ref{fig:ISO_cmd_BPass} was created from the BPASS Binary and Secular models. These models are similar to the space of the CMD regions in the PARSEC models. Although, the regions in both the Binary and Secular models span a wider color space and are shifted slightly red compared to the PARSEC models. The binary models show that stars within the 20-30 Myr age range can be around a 19th mag in F160W.  Before the BPASS binary model, the brightest stars in the 20-30 Myr age range from the results were not mirrored in the models. This provides evidence for the possibility of the three RSG candidates with age results between 20 and 30 Myr and an F160W magnitude between 19 and 19.5 as being in binary star systems. These candidates are labeled with the 'B' flag in Table \ref{tab:MATCH_results}. 

\subsection{Predicting Mass} \label{age2mass}
All of this information gathered from the models can help us predict the masses of the candidates. From the PARSEC isochrones, we can get a range of masses for each set of ages in the results. Model stars from the isochrones that pass all other cuts are allowed to be 5.4 Myr at the youngest and 26.3 at the oldest. Model stars with an age between 5.4 and 9.9 Myr will correspond to masses between 17.76 and 30.05\(M_\odot\), stars with an age between 10-19.9 Myr correspond to 11.19-17.77\(M_\odot\), and ages 20-26.3 Myr to 10-11.72\(M_\odot\), 10\(M_\odot\) being the lowest mass we were accepting values for. 

Out of the 40 highest quality candidates, four of them fall within the 17.76-30.05\(M_\odot\) mass range, 15 fall within the 11.19-17.77\(M_\odot\) range, and five fall within the 10-11.72\(M_\odot\) range. For the secondary match sample and the same mass ranges we find zero candidates, seven candidates, and one candidate, respectively. Without spectroscopy, we are unable to verify our mass constraints on these individual stars. We hope to perform such spectroscopic verification in future work. 

\section{Conclusion} \label{final}
We gathered images of NGC~6946 in the F435W, F606W, F814W, F110W, and F160W \emph{HST} filters and measured photometry using a PSF. DOLPHOT gave us quality metrics on our photometry which we used to make cuts to improve our catalogs. We applied several stellar evolutionary model sets to determine the color and luminosity space where RSG stars should appear in our photometry catalogs. After matching the \emph{HST} catalog with a Spitzer catalog, analyzing multiwavelength SEDs, comparing to Gaia, and visually inspecting \emph{HST} imgaging, we have generated a sample of stars likely to be RSGs in NGC~6946. Using MATCH we inferred ages for the candidates from their surrounding stellar populations, which we then compared to current evolutionary models and isochrones. 

Our comparisons to evolutionary models suggest that RSGs can have a broader range magnitudes and color in the Spitzer bands than the models predict. The model magnitudes and colors of RSGs in Spitzer bands are very narrow, and the data shows a much broader and bluer range of colors. On the other hand, the model predictions in the \emph{HST} IR bands more closely match those seen in the data. We found general agreement between the ages of the populations near the RSGs and the colors and luminosities predicted for RSGs at those ages. 

Furthermore, the few cases that disagreed with the ages of single-star models were better matched with the ages of binary models, making them good binary RSG candidates. From the BPASS models we can see that binary systems are a possible explanation for some of the older candidates being much brighter than the models and the other stars within the same age group. Three out of the 40 primary candidates show this evidence for being strongly affected by binary evolution.

In the future, acquiring spectroscopy of the RSG candidates identified previously would allow us to compute mass values for these stars. With profound advances in astronomical imaging on the horizon (projects such as JWST, Nancy Grace Roman Space Telescope, LSST, and Thirty Meter Telescope), studying RSG stars is not only becoming easier but is also going to be a key to understanding more about supernovae and the stellar evolution process preceding the violent death.

Support for this work was provided  by  NASA  through  grant GO-15216  from the Space Telescope Science Institute, which is operated by the Associations of  Universities  for  Research  in  Astronomy,  Incorporated,  under  NASA  contract  NAS5-26555.

A special thank you to Emily Levesque and Kathryne Neugent for advising us on our RSG candidate selection process and to look out for AGB stars.  

This work has made use of data from the European Space Agency (ESA) mission
{\it Gaia} (\url{https://www.cosmos.esa.int/gaia}), processed by the {\it Gaia}
Data Processing and Analysis Consortium (DPAC,
\url{https://www.cosmos.esa.int/web/gaia/dpac/consortium}). Funding for the DPAC
has been provided by national institutions, in particular the institutions
participating in the {\it Gaia} Multilateral Agreement.

This research has made use of the NASA/IPAC Extragalactic Database (NED), which is funded by the National Aeronautics and Space Administration and operated by the California Institute of Technology.

This research has made use of the SVO Filter Profile Service (http://svo2.cab.inta-csic.es/theory/fps/) supported from the Spanish MINECO through grant AYA2017-84089

\textit{Software:} Astrodrizzle 
\citep{Hack2012}, DOLPHOT \citep{dolphot2000,dolphot2016}, PyRAF \citep{STScI2012}, TinyTim PSFs \citep{Krist2011} MATCH \citep{Dolphin2002,Dolphin2012,Dolphin2013}

\bibliographystyle{aasjournal}
\bibliography{RSG_ref.bib}

\begin{center}
\begin{deluxetable}{cccccccc}
\tablewidth{0pt} 
\tablecolumns{8} 
\tablecaption{NGC~6946 \emph{HST} Observations. \tablenote{a}\label{tab:obs} } 
\tablehead{ \colhead{Image} & \colhead{PID} & \colhead{Date} & \colhead{RA (J2000)} & \colhead{Dec (J2000)} & \colhead{V3 Axis (deg)} & \colhead{Exp Time (sec)} & \colhead{Filter} }
\startdata
icyq02g6q & 14156 & 2016-02-09 & 308.74498 & 60.10888 & 166.93769 & 228 & WFC3 F110W \\
icyq02g8q & 14156 & 2016-02-09 & 308.74427 & 60.11086 & 166.93690 & 228 & WFC3 F110W \\
icyq02h1q & 14156 & 2016-02-09 & 308.68559 & 60.12741 & 166.88609 & 228 & WFC3 F110W \\
icyqa2h3q & 14156 & 2016-02-09 & 308.68441 & 60.12877 & 166.81579 & 228 & WFC3 F110W \\
icyqa2hdq & 14156 & 2016-02-09 & 308.62572 & 60.14583 & 166.76489 & 228 & WFC3 F110W \\
icyqa2hfq & 14156 & 2016-02-09 & 308.62489 & 60.14722 & 166.76421 & 228 & WFC3 F110W \\
icyqb2hoq & 14156 & 2016-02-09 & 308.65851 & 60.17230 & 166.74741 & 228 & WFC3 F110W \\
icyqb2hqq & 14156 & 2016-02-09 & 308.65768 & 60.17369 & 166.74659 & 228 & WFC3 F110W \\
icyqc2hzq & 14156 & 2016-02-09 & 308.71798 & 60.15385 & 166.77150 & 228 & WFC3 F110W \\
icyqc2i1q & 14156 & 2016-02-09 & 308.71715 & 60.15525 & 166.77071 & 228 & WFC3 F110W \\
icyqd2iaq & 14156 & 2016-02-09 & 308.77755 & 60.13539 & 166.78489 & 228 & WFC3 F110W \\
icyqd2icq & 14156 & 2016-02-09 & 308.77672 & 60.13678 & 166.78410 & 228 & WFC3 F110W \\
icyqd2ilq & 14156 & 2016-02-09 & 308.80985 & 60.16186 & 166.81289 & 228 & WFC3 F110W \\
icyqd2inq & 14156 & 2016-02-09 & 308.80902 & 60.16325 & 166.81210 & 228 & WFC3 F110W \\
icyqe2iwq & 14156 & 2016-02-09 & 308.75020 & 60.18035 & 166.68691 & 228 & WFC3 F110W \\
icyqe2iyq & 14156 & 2016-02-09 & 308.74936 & 60.18175 & 166.68609 & 228 & WFC3 F110W \\
icyqe2j7q & 14156 & 2016-02-09 & 308.69052 & 60.19874 & 166.63510 & 228 & WFC3 F110W \\
icyqf2j8q & 14156 & 2016-02-09 & 308.68956 & 60.20002 & 166.55819 & 228 & WFC3 F110W \\
id6o01lzq & 14638 & 2016-10-23 & 308.62762 & 60.13978 & 256.69132 & 99 & WFC3 F160W \\
id6o01m2q & 14638 & 2016-10-23 & 308.62796 & 60.13983 & 256.69159 & 99 & WFC3 F160W \\
id6o01m5q & 14638 & 2016-10-23 & 308.63039 & 60.14019 & 256.69369 & 99 & WFC3 F160W \\
id6o01m8q & 14638 & 2016-10-23 & 308.63073 & 60.14025 & 256.69400 & 99 & WFC3 F160W \\
id6o02mdq & 14638 & 2016-10-23 & 308.66598 & 60.17140 & 256.72449 & 99 & WFC3 F160W \\
... & ... & ... & ... & ... & ... & ... & .. \\
\enddata
\tablenotetext{a}{Full table available in supplemental materials.}
\end{deluxetable}
\end{center}

\begin{table}
\begin{center}
\caption{The position and photometry of our '\emph{HST}-only' sample of 17,865 stars. \label{tab:IRPhot} }
\begin{tabular}{c|c|c|c|c|c|c}
\hline
R.A. \emph{HST} & decl. \emph{HST} & F435 Mag & F606 Mag & F814 Mag & F110 Mag & F160 Mag \\ 
\hline\hline
308.5894463 &     60.1402741 &   99.999 &   26.863 &   23.966 &    21.78 &   20.587 \\
  308.5899806 &     60.1382941 &   27.899 &   25.273 &   23.521 &    22.19 &   21.116 \\
  308.5905514 &     60.1372889 &   27.558 &   24.187 &   21.612 &   20.336 &   19.192 \\
  308.5929343 &     60.1384478 &   30.041 &   29.846 &   27.897 &   21.948 &   20.923 \\
  308.5929727 &     60.1384630 &   29.023 &   25.228 &   22.378 &   20.953 &   19.669 \\
  308.5950922 &     60.1438573 &   27.636 &   24.567 &   22.389 &   21.011 &   20.033 \\
  308.5959332 &     60.1409364 &   27.805 &   25.228 &   23.292 &   21.928 &   20.751 \\
  308.5959735 &     60.1381955 &   27.919 &   24.415 &   21.999 &   20.627 &   19.406 \\
  308.5961989 &     60.1434902 &   27.402 &    24.45 &   21.936 &   20.577 &   19.547 \\
  308.5970938 &     60.1439234 &   29.888 &   30.468 &   25.676 &   21.431 &   20.437 \\
  308.5972989 &     60.1373564 &   29.371 &   25.489 &     23.4 &   22.192 &   21.102 \\
  308.5979507 &     60.1369945 &   28.065 &   24.596 &   22.064 &   20.474 &   19.424 \\
  308.5983302 &     60.1453626 &   33.848 &   25.745 &   23.051 &   21.399 &    20.37 \\
  308.5987795 &     60.1460313 &   27.638 &   24.909 &   22.505 &   20.655 &   19.659 \\
  308.5993380 &     60.1341256 &   29.136 &   26.388 &   23.648 &   22.238 &   21.243 \\
  308.5994366 &     60.1479948 &   27.704 &   25.245 &   22.834 &   21.217 &   20.169 \\
  308.5998366 &     60.1400633 &   28.743 &   25.593 &   23.718 &   22.378 &   21.254 \\
  308.6005390 &     60.1476767 &   28.987 &   25.055 &   23.019 &   21.538 &   20.443 \\
  308.6013874 &     60.1435820 &   27.237 &   24.318 &   22.468 &   21.451 &    20.37 \\
  308.6014281 &     60.1422371 &    27.53 &    24.89 &   23.243 &   22.257 &   21.195 \\
  308.6015950 &     60.1505382 &   99.999 &   99.999 &   31.286 &   22.111 &   21.156 \\
  308.6016186 &     60.1505501 &   99.999 &   26.409 &   22.775 &   20.841 &   19.864 \\
  308.6019077 &     60.1512509 &   27.748 &   24.915 &    23.09 &    21.91 &   20.922 \\
  308.6019790 &     60.1517156 &   27.678 &   25.475 &    23.59 &   22.201 &   21.108 \\
  308.6020586 &     60.1499165 &   28.771 &   25.879 &   23.487 &   21.867 &   20.645 \\
  308.6020997 &     60.1509134 &   28.896 &   25.461 &    23.23 &   21.825 &   20.732 \\
  308.6022433 &     60.1373140 &   26.988 &   24.586 &   22.699 &   21.509 &   20.464 \\
  308.6023214 &     60.1519305 &   27.918 &    25.08 &   23.437 &   22.233 &   21.265 \\
  308.6026652 &     60.1501214 &   30.906 &   26.065 &   23.572 &   21.739 &   20.558 \\
  308.6028151 &     60.1519660 &   27.654 &   25.427 &   23.172 &   21.824 &   20.664 \\
  308.6028420 &     60.1508512 &   29.031 &   26.358 &   23.094 &   20.771 &   19.805 \\
  308.6029213 &     60.1518894 &   28.103 &   25.492 &    23.63 &   22.313 &   21.292 \\
  308.6030309 &     60.1506125 &   30.459 &   25.682 &   23.426 &   21.903 &   20.912 \\
  308.6031192 &     60.1504433 &   25.253 &   24.297 &   22.853 &   21.741 &    20.72 \\
  308.6036771 &     60.1456030 &   29.304 &   28.308 &   25.647 &   21.663 &   20.442 \\
  308.6037083 &     60.1518668 &   25.753 &   24.647 &   23.244 &   22.073 &   21.035 \\
  308.6038038 &     60.1521061 &   23.609 &   23.126 &    22.35 &   21.603 &   20.603 \\
  308.6038821 &     60.1520461 &   28.208 &   24.646 &   22.614 &   21.474 &   20.427 \\
  308.6039796 &     60.1521532 &   28.318 &   25.602 &   23.348 &   21.876 &   20.834 \\
  308.6040707 &     60.1445576 &   27.716 &   25.154 &   23.283 &   21.941 &   20.931 \\
  ... & ... & ... & ... & ... & ... & ... \\
\hline
\end{tabular}
\tablenotetext{a}{The first two columns are the R.A. and decl. in the \emph{HST} catalogs (J200). The remaining columns are magnitudes in the F435, F606, F814, F110, F160 filters. A value of 99 indicates that the star was not recovered in the filter.}
\end{center}
\end{table}

\begin{sidewaystable}
\centering
\caption{Photometry for each candidate.}
\begin{tabular}{c|c|c|c|c|c|c|c|c|c|c|c|c}
\label{tab:Phot}
Identifier & R.A. \emph{HST} & decl. \emph{HST} & R.A. \emph{Spitzer} & decl. \emph{Spitzer} & F435 Mag & F606 Mag & F814 Mag & F110 Mag & F160 Mag & 3.6m Mag & 4.5m Mag & 5.8m Mag \\
\hline
      RSG 2669 &   308.6038821 &     60.1520461 &         308.60391 &           60.15193 &   28.208 &   24.646 &   22.614 &   21.474 &   20.427 &    16.94 &    17.05 &    15.55 \\
      RSG 2055 &   308.6066464 &     60.1462848 &         308.60663 &           60.14621 &   23.972 &   23.081 &    22.27 &    21.36 &   20.354 &    16.64 &    16.69 &     15.9 \\
      RSG 6602 &   308.6109976 &     60.1382102 &         308.61094 &           60.13817 &   27.052 &   24.227 &   22.282 &   21.176 &   20.087 &    18.69 &    18.83 &    16.09 \\
      RSG 1851 &   308.6215717 &     60.1405858 &         308.62158 &           60.14064 &   27.865 &   25.017 &    23.35 &   22.131 &   21.211 &    16.53 &    16.56 &    14.15 \\
      RSG 6318 &   308.6232846 &     60.1269144 &         308.62319 &           60.12691 &   27.666 &   25.533 &   23.223 &   21.522 &   20.498 &    18.36 &    18.67 &    15.87 \\
      RSG 1278 &   308.6236309 &     60.1362338 &         308.62365 &           60.13613 &   26.261 &   24.765 &   23.578 &    22.12 &   21.119 &    16.03 &     16.1 &    13.65 \\
      RSG 3066 &   308.6260371 &     60.1392559 &         308.62595 &           60.13932 &   99.999 &   25.754 &    23.37 &   21.662 &   20.616 &    16.72 &    17.22 &    15.33 \\
      RSG 1356 &   308.6265974 &     60.1384984 &         308.62651 &           60.13855 &   25.909 &   25.245 &   22.927 &   20.987 &   19.795 &    16.14 &    16.17 &    14.62 \\
       RSG 650 &   308.6276566 &     60.1396612 &         308.62766 &            60.1396 &    26.25 &   24.122 &   22.405 &   20.822 &   19.756 &    15.31 &    15.33 &    14.95 \\
      RSG 2214 &   308.6276590 &     60.1714820 &         308.62753 &           60.17153 &   27.506 &   25.941 &   99.999 &   22.321 &   21.252 &    16.72 &    16.79 &    13.93 \\
      RSG 6822 &   308.6299857 &     60.1595435 &         308.62993 &           60.15951 &   99.999 &   27.681 &   99.999 &   21.603 &   20.413 &    18.39 &    18.96 &    16.19 \\
       RSG 356 &   308.6301388 &     60.1380795 &         308.63021 &           60.13805 &   26.809 &   24.728 &   22.753 &   21.396 &   20.433 &    14.53 &    14.78 &    12.95 \\
      RSG 1198 &   308.6302907 &     60.1363369 &         308.63021 &           60.13636 &   26.842 &   25.538 &   23.218 &   21.588 &   20.579 &    15.97 &    16.02 &    13.93 \\
      RSG 4741 &   308.6310596 &     60.1627529 &          308.6312 &           60.16278 &   99.999 &   29.775 &   99.999 &   22.299 &   21.201 &    17.85 &    17.95 &    16.01 \\
      RSG 4906 &   308.6323098 &     60.1629045 &         308.63247 &           60.16302 &   26.637 &   24.644 &   99.999 &   21.002 &   19.956 &    18.01 &    18.02 &    14.94 \\
        RSG 56 &   308.6333372 &     60.1374750 &         308.63337 &            60.1375 &   23.735 &   23.141 &   21.833 &   20.203 &   19.035 &    12.66 &    12.94 &    12.35 \\
       RSG 347 &   308.6347831 &     60.1737147 &         308.63462 &           60.17361 &   99.999 &   25.301 &   99.999 &    21.41 &   20.289 &    14.36 &    14.76 &     11.5 \\
      RSG 1764 &   308.6372782 &     60.1342722 &         308.63729 &           60.13433 &   25.069 &    24.26 &   23.132 &   22.058 &   21.127 &    16.37 &    16.51 &    13.49 \\
      RSG 1785 &   308.6394347 &     60.1697384 &         308.63941 &           60.16976 &   30.238 &   26.023 &   99.999 &   22.074 &   20.921 &    16.43 &    16.52 &    14.07 \\
      RSG 4325 &   308.6397297 &     60.1417534 &         308.63958 &           60.14182 &    28.41 &   26.563 &    23.79 &   21.884 &   20.917 &    17.72 &    17.77 &    15.99 \\
      RSG 6222 &   308.6423098 &     60.1252526 &         308.64223 &           60.12535 &   26.323 &   25.354 &   24.142 &   22.272 &   21.173 &    18.42 &    18.62 &    15.74 \\
      RSG 4825 &   308.6475803 &     60.1875352 &         308.64769 &           60.18753 &   99.999 &   99.999 &   99.999 &   21.496 &   20.286 &    17.76 &    17.98 &    15.43 \\
      RSG 3223 &   308.6481818 &     60.1647505 &         308.64838 &           60.16468 &   27.374 &   25.124 &   99.999 &   21.832 &   20.886 &    16.84 &     17.3 &    16.73 \\
      RSG 5027 &   308.6495598 &     60.1496295 &         308.64945 &           60.14959 &   27.461 &   24.879 &   22.121 &   20.256 &   19.253 &    17.87 &    18.07 &    15.32 \\
      RSG 2398 &   308.6497780 &     60.1507247 &         308.64983 &            60.1508 &   27.161 &    25.38 &   23.187 &   21.587 &   20.625 &    16.69 &     16.9 &    14.48 \\
      RSG 5026 &   308.6503162 &     60.1593026 &         308.65039 &           60.15925 &    27.59 &   23.832 &   99.999 &    20.16 &   19.066 &    17.62 &    18.07 &     13.2 \\
      RSG 5542 &   308.6509015 &     60.1479127 &         308.65078 &           60.14791 &   99.999 &   26.307 &    23.14 &   21.202 &   20.211 &    18.11 &    18.29 &    15.43 \\
      RSG 3282 &   308.6520437 &     60.1569410 &         308.65216 &           60.15698 &   26.302 &   25.051 &   23.278 &   21.522 &   20.433 &    17.23 &    17.33 &    15.29 \\
      RSG 1590 &   308.6527846 &     60.1347001 &         308.65278 &           60.13482 &   23.961 &   23.229 &   22.571 &   22.012 &   20.938 &    16.37 &    16.38 &    15.32 \\
      RSG 4203 &   308.6530159 &     60.1548841 &         308.65293 &           60.15484 &   27.149 &   25.168 &   23.767 &   22.081 &   21.013 &     17.4 &    17.73 &    16.07 \\
      RSG 3991 &   308.6559471 &     60.1528615 &           308.656 &           60.15275 &   27.775 &   24.711 &   22.267 &   20.574 &   19.533 &    17.19 &    17.64 &     15.7 \\
      RSG 3984 &   308.6568806 &     60.1878111 &         308.65693 &           60.18783 &   99.999 &   99.999 &   99.999 &   22.364 &   21.272 &    17.51 &    17.63 &    15.25 \\
      RSG 2813 &   308.6584383 &     60.1221435 &         308.65861 &           60.12224 &   27.088 &   24.549 &   22.077 &   20.536 &   19.385 &    16.85 &    17.11 &    15.12 \\
      RSG 1900 &   308.6591681 &     60.1535855 &         308.65904 &           60.15366 &   27.181 &   25.287 &   99.999 &   22.119 &   21.083 &    16.47 &    16.58 &    14.41 \\
      RSG 5871 &   308.6619373 &     60.1434257 &          308.6618 &            60.1434 &   28.013 &    24.83 &   99.999 &   20.933 &   19.994 &    17.94 &    18.44 &    15.99 \\
      RSG 2666 &   308.6640101 &     60.1857048 &         308.66402 &           60.18568 &   99.999 &   25.356 &   22.641 &   20.872 &   19.893 &    16.49 &    17.05 &    16.83 \\
       RSG 422 &   308.6671856 &     60.1723374 &         308.66723 &           60.17225 &   26.765 &   25.488 &   24.157 &   22.125 &   21.221 &    14.74 &    14.91 &    14.78 \\
      RSG 2788 &   308.6708826 &     60.1573541 &         308.67101 &           60.15728 &   24.868 &   23.872 &   22.866 &   22.024 &   20.934 &    16.79 &     17.1 &    14.76 \\
      RSG 2965 &   308.6717680 &     60.1513484 &         308.67181 &           60.15131 &   25.061 &   23.883 &   99.999 &   21.887 &   20.905 &    17.07 &    17.18 &    15.44 \\
... & ... & ... & ... & ... & ... & ... & ... & ... & ... & ... & ... \\
\hline
\end{tabular}
\tablenotetext{a}{A value of 99 indicates that the star was not recovered in the filter.}
\end{sidewaystable}

\newpage

\begin{center}
\begin{longtable}{cccccc}
\caption{Full list of results from MATCH. \label{tab:MATCH_results}} \\ 
\hline
\hline
Identifier & $dA_{v}$ & $A_{v}$ & Age (Myr) & Flags \\
\hline
\endfirsthead
\hline
\multicolumn{5}{c}{Continuation of Table \ref{tab:MATCH_results}}\\
\hline
Identifier & $dA_{v}$ & $A_{v}$ & Age (Myr) & Flags \\
\hline
\endhead
\hline
\endfoot
\hline
\endlastfoot
RSG 2669 &  0.2 &  1.15 &   16 &       M,2 \\
 RSG 2055 &    2 &  2.35 &    4 &       F,2 \\
 RSG 6602 &  0.2 &   1.4 &    0 &       C,1 \\
 RSG 1851 &    0 &   1.5 &   11 &       M,2 \\
 RSG 6318 &    0 &  1.65 &   18 &       A,1 \\
 RSG 1278 &    0 &   1.2 &   26 &       F,1 \\
 RSG 3066 &  0.2 &   1.6 &   16 &       M,2 \\
 RSG 1356 &  0.4 &   1.5 &   13 &       A,2 \\
  RSG 650 &  0.8 &   1.6 &    5 &       M,2 \\
 RSG 2214 &  0.2 &  1.35 &   27 &     A,D,1 \\
 RSG 6822 &    0 &   1.5 &   12 &     A,D,1 \\
  RSG 356 &  0.4 &   0.9 &   13 &       M,1 \\
 RSG 1198 &  0.2 &  1.05 &   13 &       M,2 \\
 RSG 4741 &  N/A &   N/A &  N/A &  D,M,IN,1 \\
 RSG 4906 &    0 &  1.05 &   11 &     A,D,1 \\
   RSG 56 &    1 &     1 &    9 &       F,2 \\
  RSG 347 &  1.6 &  1.35 &   29 &     D,F,1 \\
 RSG 1764 &    0 &  1.05 &    4 &       M,1 \\
 RSG 1785 &  0.4 &  1.75 &    0 &     C,D,1 \\
 RSG 4325 &    0 &  1.55 &   19 &       M,1 \\
 RSG 6222 &  0.2 &  1.55 &    0 &       C,1 \\
 RSG 4825 &  N/A &   N/A &  N/A &    D,IN,1 \\
 RSG 3223 &  1.2 &  1.25 &   29 &     D,M,1 \\
 RSG 5027 &    0 &  1.25 &    0 &       C,1 \\
 RSG 2398 &    0 &  1.85 &   16 &       F,1 \\
 RSG 5026 &    0 &  1.15 &   29 &     A,D,1 \\
 RSG 5542 &    0 &   1.5 &   16 &       A,2 \\
 RSG 3282 &    0 &  1.85 &   21 &       A,1 \\
 RSG 1590 &    0 &   2.1 &    7 &       F,2 \\
 RSG 4203 &    0 &   1.4 &   23 &       M,2 \\
 RSG 3991 &  1.2 &   1.5 &    5 &       A,1 \\
 RSG 3984 &  N/A &   N/A &  N/A &    D,IN,1 \\
 RSG 2813 &    0 &  1.15 &   26 &       A,1 \\
 RSG 1900 &  0.4 &   1.5 &    4 &     D,M,2 \\
 RSG 5871 &    0 &   1.5 &    9 &     D,M,2 \\
 RSG 2666 &    0 &   1.2 &   15 &       A,2 \\
  RSG 422 &  0.8 &   1.6 &    0 &     C,F,1 \\
 RSG 2788 &  1.2 &  2.15 &    6 &       F,1 \\
 RSG 2965 &  0.4 &   1.9 &   27 &     D,F,1 \\
 RSG 1839 &    0 &   1.4 &   15 &       M,1 \\
 RSG 6480 &  0.4 &   1.1 &   11 &       M,2 \\
 RSG 3105 &    0 &   1.5 &   17 &     A,D,2 \\
 RSG 4781 &    0 &  1.35 &    5 &       A,1 \\
 RSG 5914 &  N/A &   N/A &  N/A &  D,M,IN,1 \\
 RSG 2768 &  0.4 &  2.05 &   23 &       A,1 \\
 RSG 2841 &    0 &     1 &    4 &       A,1 \\
 RSG 4393 &  1.2 &  1.05 &   13 &     A,D,2 \\
 RSG 1282 &    0 &   1.3 &   14 &       A,1 \\
  RSG 477 &    0 &     1 &    8 &       F,1 \\
 RSG 2132 &    0 &  2.15 &   15 &     D,M,2 \\
 RSG 1966 &    0 &   1.5 &    5 &       A,1 \\
 RSG 1321 &    0 &   1.6 &   15 &       A,1 \\
 RSG 3485 &    0 &   1.3 &    9 &     A,D,1 \\
 RSG 2171 &  0.2 &   1.8 &   16 &       A,1 \\
 RSG 1132 &    0 &   1.6 &   23 &       M,2 \\
 RSG 3129 &    0 &   1.5 &   13 &       M,2 \\
 RSG 2464 &    0 &   1.7 &    5 &       M,2 \\
 RSG 3160 &  0.2 &   0.9 &   16 &       A,1 \\
 RSG 2556 &    0 &  1.85 &   13 &       M,2 \\
 RSG 4972 &    0 &   1.4 &   16 &       A,1 \\
  RSG 611 &  1.6 &   1.6 &    4 &     D,F,1 \\
 RSG 3759 &  1.2 &   1.8 &   27 &       A,2 \\
 RSG 3532 &    0 &  1.55 &   29 &     A,D,1 \\
 RSG 3059 &  2.2 &  1.75 &   15 &     A,D,1 \\
 RSG 3561 &    0 &  1.15 &   16 &       A,1 \\
 RSG 4502 &    0 &  1.35 &   29 &       A,1 \\
  RSG 997 &  1.2 &   1.1 &   17 &       F,2 \\
  RSG 814 &  N/A &   N/A &  N/A &    D,IN,1 \\
  RSG 959 &    0 &   1.1 &    7 &       M,1 \\
  RSG 350 &    0 &   1.6 &    5 &       M,1 \\
 RSG 2218 &  0.4 &  1.15 &    4 &       M,2 \\
 RSG 2748 &    0 &     2 &   21 &     A,D,1 \\
  RSG 578 &  0.8 &   0.8 &   15 &       F,1 \\
  RSG 309 &  2.6 &   2.3 &    4 &       F,2 \\
 RSG 1465 &  N/A &   N/A &  N/A &  D,F,IN,2 \\
  RSG 759 &  0.8 &   0.9 &   23 &       M,2 \\
 RSG 4176 &    0 &  1.15 &   16 &       M,2 \\
 RSG 2424 &    0 &  1.15 &   14 &       A,1 \\
   RSG 51 &  N/A &   N/A &  N/A &  D,F,IN,1 \\
 RSG 1622 &    0 &   1.8 &    6 &       M,1 \\
  RSG 798 &  0.2 &  1.15 &    5 &     D,F,1 \\
 RSG 1355 &  0.2 &  0.85 &   10 &     A,D,1 \\
 RSG 1661 &  0.4 &   1.7 &    5 &       A,1 \\
  RSG 928 &    0 &  1.95 &   24 &     A,D,1 \\
  RSG 542 &  N/A &   N/A &  N/A &  D,M,IN,1 \\
 RSG 1592 &    0 &  1.25 &    4 &     A,D,2 \\
 RSG 5460 &    0 &  1.35 &   16 &       A,1 \\
 RSG 2722 &    0 &  1.55 &   21 &       M,1 \\
 RSG 3935 &  1.4 &  1.05 &   24 &     D,M,1 \\
  RSG 383 &  0.4 &  1.35 &   24 &     D,F,1 \\
 RSG 2006 &    0 &   1.3 &   19 &       M,2 \\
  RSG 935 &    0 &  1.45 &   17 &     D,M,1 \\
 RSG 2226 &    0 &  1.25 &    8 &     D,F,1 \\
 RSG 1385 &    0 &  1.65 &    6 &       M,1 \\
 RSG 5007 &    0 &  1.35 &   10 &       M,1 \\
 RSG 1539 &  0.2 &   1.6 &    5 &       A,1 \\
 RSG 2680 &  0.2 &   0.8 &    4 &       A,2 \\
 RSG 3158 &    0 &  0.95 &    0 &       C,1 \\
  RSG 527 &  0.8 &   1.4 &    4 &     D,M,1 \\
 RSG 2141 &    0 &  1.45 &    5 &     D,M,2 \\
  RSG 734 &    0 &  1.45 &    5 &     D,F,1 \\
 RSG 3954 &    0 &     1 &   10 &       A,1 \\
 RSG 4092 &  0.4 &   1.6 &   13 &       F,2 \\
 RSG 1919 &  1.6 &  1.05 &    5 &     D,M,2 \\
 RSG 3091 &    0 &  1.65 &    6 &       A,1 \\
 RSG 1489 &    0 &   1.4 &    0 &       C,1 \\
 RSG 4665 &  N/A &   N/A &  N/A &    D,IN,1 \\
 RSG 4480 &  0.2 &   1.5 &   26 &       A,1 \\
 RSG 2276 &    0 &  1.55 &   21 &       A,2 \\
 RSG 4966 &  N/A &   N/A &  N/A &    D,IN,1 \\
 RSG 5120 &  1.4 &  1.15 &   14 &       A,2 \\
 RSG 4949 &    0 &   1.6 &    9 &       A,1 \\
 RSG 5381 &    1 &   2.1 &    6 &     D,M,1 \\
  RSG 694 &  0.2 &   1.5 &   13 &     D,M,2 \\
  RSG 677 &  1.8 &   2.6 &   24 &     D,F,1 \\
 RSG 1534 &  1.6 &  1.45 &    4 &       A,1 \\
 RSG 4118 &  0.2 &   2.1 &   29 &       A,1 \\
  RSG 637 &  0.2 &   0.9 &    4 &     D,F,1 \\
 RSG 4489 &  1.2 &   1.8 &    0 &       C,1 \\
 RSG 3640 &  N/A &   N/A &  N/A &    D,IN,1 \\
  RSG 692 &    0 &  1.05 &    6 &     D,F,2 \\
  RSG 447 &    0 &   1.3 &    4 &     D,F,1 \\
 RSG 3441 &  N/A &   N/A &  N/A &    D,IN,1 \\
 RSG 1930 &    0 &   1.3 &    0 &     C,M,1 \\
 RSG 3264 &    0 &  1.35 &   11 &     D,F,1 \\
 RSG 4623 &    0 &  1.45 &   29 &     A,D,1 \\
 RSG 1713 &  0.2 &   1.8 &   26 &       F,1 \\
 RSG 3199 &    0 &   1.5 &   14 &     D,M,2 \\
  RSG 767 &  0.4 &  1.65 &    9 &     A,D,1 \\
 RSG 1480 &    0 &   1.3 &    4 &     D,F,1 \\
 RSG 1189 &    0 &   1.6 &    4 &     D,F,1 \\
 RSG 4881 &    0 &  1.55 &   27 &     A,D,1 \\
 RSG 1744 &  0.2 &   1.1 &   19 &       F,2 \\
 RSG 3335 &    0 &   1.1 &   13 &     A,D,1 \\
 RSG 3163 &    0 &  1.15 &   21 &     D,F,2 \\
 RSG 3336 &  1.2 &   0.8 &   23 &       A,1 \\
 RSG 3113 &    0 &  1.05 &   16 &       A,2 \\
\end{longtable}
\tablenotetext{a}{Column 5 is a flag indicating the status of a candidate. A means the candidate has shown only signs of being a quality candidate. B indicates the candidate is in a potential binary system, C means the star is a potential low mass contaminant (AGB star), D means the candidate location was analyzed with the F435W and F606W filter combination, F means the star is a potential foreground star, M indicates the candidates that were our most likely mismatches, and IN means the location had insufficient coverage in the surrounding area to get an age estimate from MATCH. A flag of 1 means the candidate belongs to the primary sample comprised of only single matches between \emph{HST} and Spitzer sources while 2 means the candidate belongs to the secondary sample of candidates that had multiple matches between \emph{HST} and Spitzer sources.}
\end{center}

\newpage

\begin{figure}
\label{fig:fset}
\figurenum{16}
\plotone{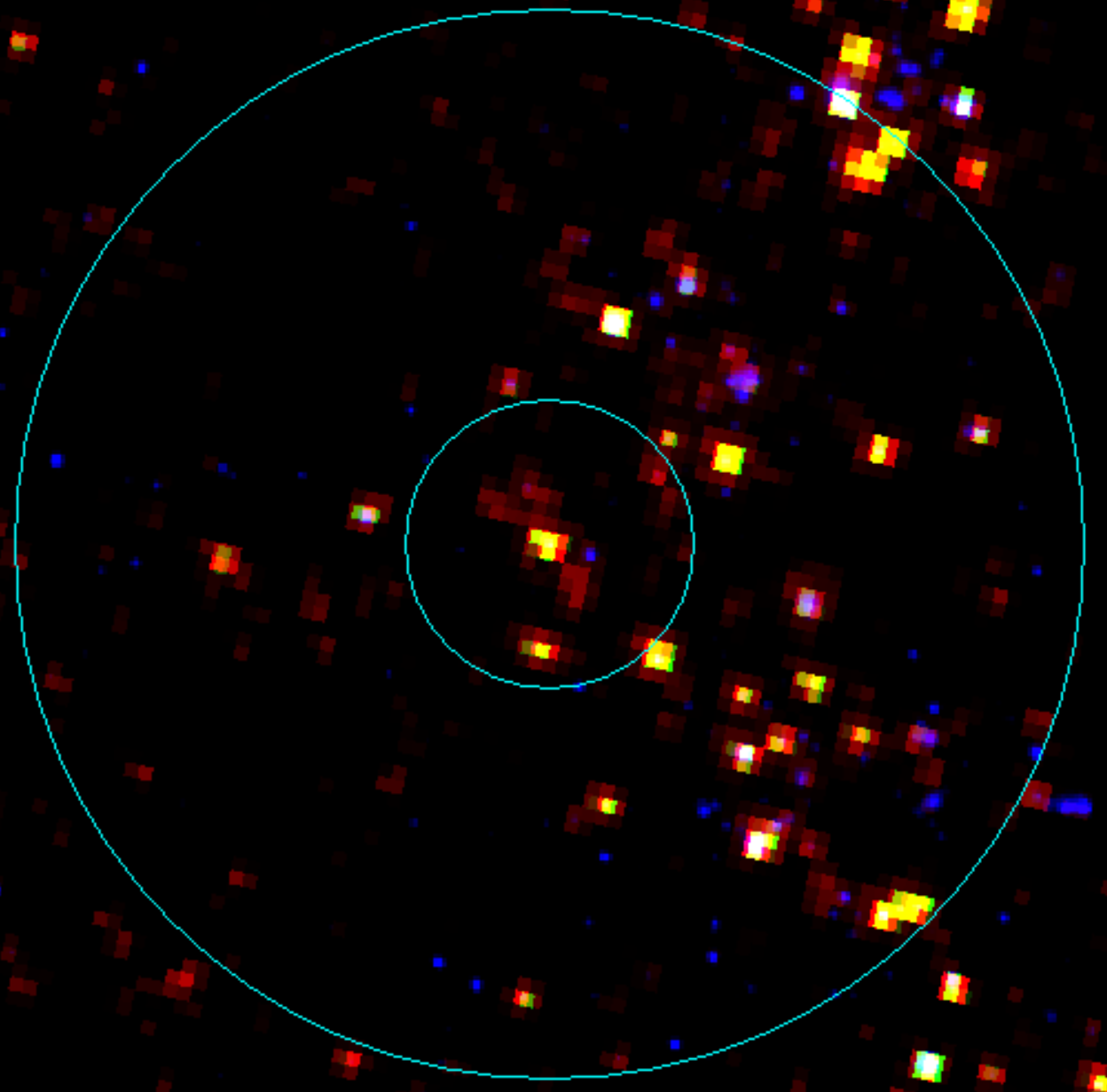}
\caption{Remaining 31 RGB Snapshots of 'A' labeled stars in the primary sample are available in the online journal.}
\end{figure}


 \begin{figure}[!htb]
    \setcounter{figure}{16}
    \center{\includegraphics[width=\textwidth]{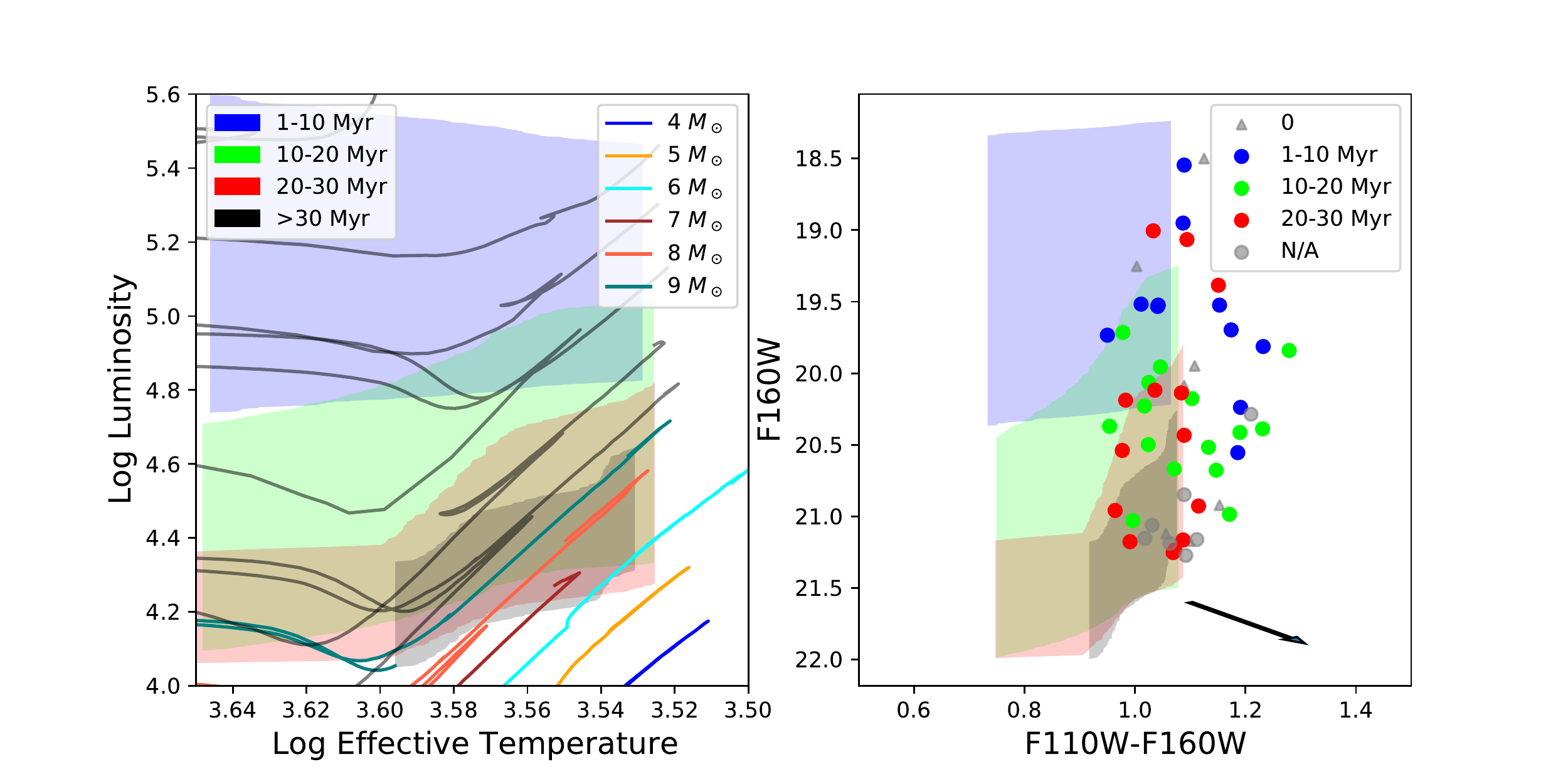}}
    \caption{\label{fig:possAGB} Similar to Figure \ref{fig:ISO_cmd} except with lower-mass stars plotted on the left panel. The effective temperature and luminosity limits are also removed. This shows where possible AGB stars could mix with our selection of RSG stars. The tracks of the lower-mass stars have lower masses than we previously allowed in the models and are older than the maximum age allowed in our results from MATCH. We see that some of the fainter stars in our sample are potential AGB contaminants.}
 \end{figure} 

\end{document}